\documentclass[superscriptaddress,
 aps, prl, twocolumn, 10pt,
 floatfix, amsmath, amssymb, notitlepage,
]{revtex4-2}

\usepackage[utf8]{inputenc}
\usepackage{parskip}
\usepackage{braket}
\usepackage{graphicx}
\usepackage{tabularx}
\usepackage{dcolumn}
\usepackage{booktabs}
\usepackage{amsmath,amssymb}
\usepackage{bm}
\usepackage{nicefrac}
\usepackage[normalem]{ulem}
\usepackage[dvipsnames]{xcolor}
\usepackage{placeins}\usepackage[colorlinks,linkcolor=blue,citecolor=blue,urlcolor=blue]{hyperref}
\begin{document}

\title{Designing Flat Bands, Localized and Itinerant States in TaS$_2$ Trilayer Heterostructures}

\author{Hyeonhu Bae}
\affiliation{Department of Condensed Matter Physics, Weizmann Institute of Science, 7610001 Rehovot, Israel}
\author{Roser Valent\'i}
\affiliation{Institute f{\"u}r Theoretische Physik, Goethe-Universit{\"a}t Frankfurt, D-60438 Frankfurt am Main, Germany}
\author{Igor I. Mazin}
\email{imazin2@gmu.edu}
\affiliation{Department of Physics and Astronomy, George Mason University, Fairfax, VA 22030}
\affiliation{Quantum Science and Engineering Center, George Mason University, Fairfax, VA 22030}
\author{Binghai Yan}
\email{binghai.yan@weizmann.ac.il}
\affiliation{Department of Condensed Matter Physics, Weizmann Institute of Science, 7610001 Rehovot, Israel}
\affiliation{Department of Physics, the Pennsylvania State University, University Park, PA, 16802, USA }
\date{\today}

\begin{abstract}
{Stacking and twisting van der Waals materials provide a powerful tool to design quantum matter and engineer electron correlation. For instance, monolayers of 1T- and 1H-TaS$_2$ are Mott insulating and metallic (also superconducting), respectively, and thus, the T/H bilayer systems have been extensively investigated in the context of heavy fermions and unconventional superconductivity, which are expected phases from localized spins (1T) coexisting with itinerant electrons (1H). However, recent studies revealed that significant charge transfer from the 1T to 1H layers removes the 1T Mottness and renders the above scenario elusive. In this work, we propose a T/T/H trilayer heterostructure by combining a T/T bilayer -- which is a band insulator with flat dispersion -- with a 1H layer. After charge redistribution, this trilayer heterostructure shows localized spins in the Mott flat band of the T/T bilayer and weak spin polarization in the metallic H layer. We argue that by varying the stacking configurations of the T/T bilayer in the T/T/H trilayer, a crossover from a doped Mott insulator to a Kondo insulator can be achieved. The T/T/H trilayer provides therefore a rich novel heterostructure platform to study strong correlation phenomena and unconventional superconductivity.}
\end{abstract}
\maketitle

\textit{Introduction} --
Van der Waals (vdW) materials have recently revolutionized the study of quantum matter, providing unprecedented control over electronic correlations and emergent phases~\cite{geim2013van,novoselov20162d,balents2020superconductivity}. A prime example is twisted bilayer graphene, where flat bands amplify Coulomb interactions, yielding unconventional superconductivity and correlated insulators~\cite{bistritzer2011moire,balents2020superconductivity,cao2018unconventional,jiang2019charge}. Inspired by such discoveries, transition metal dichalcogenides (TMDCs) have emerged as versatile platforms for engineering quantum phases~\cite{wilson2001charge,yang2017structural}.
Particularly TaX$_2$ (X=S, Se) have been highlighted due to its distinctive properties in various polytypes. Among the polytypes of TaX$_2$, 1T-TaX$_2$ is a Mott insulator with $\sqrt{13}\times\sqrt{13}$ charge density wave (CDW) structure forming Star-of-David (SoD) clusters~\cite{wilson2001charge,wang2020band,chen2020strong,nakata2021robust}. It forms a spin-1/2 triangular superlattice that is being intensively discussed as a quantum spin liquid candidate~\cite{law20171t,he2018spinon,ribak2017gapless,klanjvsek2017high,murayama2020effect,manas2021quantum,benedivcivc2020superconductivity,ruan2021evidence,chen2022evidence}. In contrast, 1H-TaX$_2$ is an Ising superconductor with a competing $3\times3$ CDW phase~\cite{coleman1983dimensional,de2018tuning,bhoi2016interplay,lian2019coexistence}.

Heterostructures made of the combination of these contrasting states, as it is the case of 4H$_\textrm{b}$-TaS$_2$, which consists of alternating 1T and 1H layers, have been exploited as promising platforms for exploring emergent phenomena resulting from the coexistence of Mott insulating and superconducting phases, including unconventional superconductivity~\cite{ribak2020chiral,nayak2021evidence,persky2022magnetic,yan2023modulating} and Kondo-like behavior~\cite{vavno2021artificial,ayani2024probing,ruan2021evidence,chen2022evidence}. However, in such heterostructures, a considerable charge transfer from the 1T to the 1H layer is expected, due to, among others, the different work functions of the layers. The 1H layer takes up to one electron per SoD away from the 1T layer in 4H$_\textrm{b}$-TaS$_2$ and in bilayer 1T/1H-TaS$_2$, depending on the distance between layers. This complicates the preservation of a Mott insulating half-filled spin-1/2 triangular lattice at the 1T layer hybridizing with the itinerant electrons of the 1H layer, and leads instead to a heavily doped Mott system~\cite{crippa2024heavy,almoalem2024charge}. These challenges have led to controversial interpretations of sample-dependent electronic properties in these platforms, as some samples seem to display Kondo-like behavior due to localized spins hybridizing with itinerant electrons~\cite{vavno2021artificial,ayani2024probing}, while other samples do not observe localized spins as a result of the strong charge transfer that vacates the localized state of 1T SoD~\cite{nayak2021evidence,wen2021roles,kumar2023first}.

In this work, we propose an alternative heterostructure, a trilayer T/T/H-TaS$_2$, which promises to display a rich variety of correlated phases and yet is free from the difficulties in T/H bilayer. While the T/T bilayer is a band insulator, where the interlayer hybridization quenches the spin degrees of freedom and disrupts the Mott bands~\cite{ritschel2018stacking,darancet2014three,butler2020mottness}, the nearly one electron transfer to the 1H monolayer leaves behind a single electron in the T/T bilayer, restoring Mott behavior. The interlayer stacking of the T/T bilayer controls the hybridization strength and charge transfer and indicates the possibility in the trilayer of a crossover from a doped Mott insulator (similar to the T/H bilayer case) to a Kondo insulator. Apart from the presence of localized spins in the T/T bilayer, the 1H monolayer exhibits spin-polarized itinerant electrons and may lead to unconventional superconductivity. In addition, we note that some recent experiments demonstrate the feasibility of realizing such trilayer heterostructures by thermal annealing~\cite{wang2018surface,sung2022two,sung2024endotaxial,husremovic2023encoding}, and laser-induced polytype transformation~\cite{ravnik2019strain,ravnik2021quantum} in bulk 1T-TaS$_2$.

\textit{Computational Methods} --
Spin-polarized density functional theory (DFT) calculations were performed using the Vienna \textit{ab initio} Simulation Package (VASP) implementing the projector augmented wave method~\cite{kresse1999ultrasoft}. We used the Perdew-Burke-Ernzerhof (PBE) exchange-correlation energy functional~\cite{perdew1996generalized}, and incorporated van der Waals interactions using the DFT-D3 method with the Becke-Johnson damping function~\cite{grimme2011effect}. The electronic correlation in Ta 5d electrons in 1H-TaS$_2$ and 1T-TaS$_2$ layers is modeled within the DFT+U implementation of Dudarev \textit{et al.}~\cite{dudarev1998electron} and considering $U_\textrm{eff,1H}$=2.82 eV and $U_\textrm{eff,1T}$=1.76 eV~\cite{ayani2024unveiling}, respectively. The kinetic energy cutoff was set to 400 eV for the plane-wave basis. The first Brillouin zone is sampled by $8\times8$ $\Gamma$-centered k-point mesh.

\begin{figure}
    \centering
    \includegraphics[width=0.9\columnwidth]{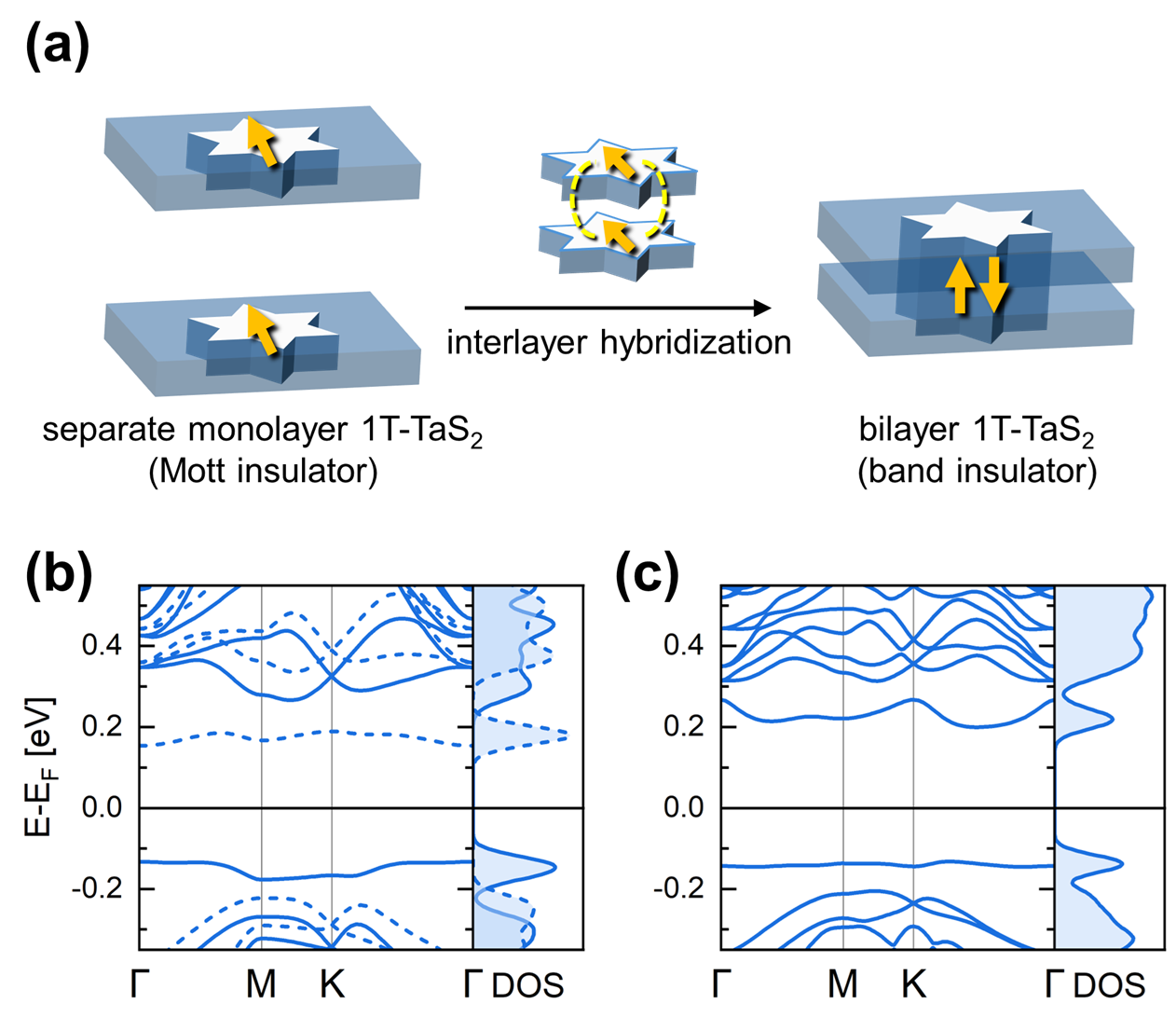}
    \caption{\textbf{Electronic structure of monolayer 1T-TaS$_2$ and bilayer T/T-TaS$_2$.}
    (\textbf{a}) Schematic illustration showing the transition from a Mott insulator (monolayer) to a band insulator (bilayer) driven by interlayer hybridization. The yellow arrows represent local spins at the center of the Star of David which get quenched in the band insulator. (\textbf{b,c}) corresponding band structures and density of states for (b) monolayer 1T-TaS$_2$ and (c) bilayer T/T-TaS$_2$, demonstrating the evolution from a correlation-driven gap to a hybridization gap. Majority and minority spins are denoted as a solid and dashed lines, respectively. The Fermi level is set to 0 eV.
    }
    \label{fig:fig1-str}
\end{figure}

\begin{figure}[tb]
    \centering
    \includegraphics[width=1.0\columnwidth]{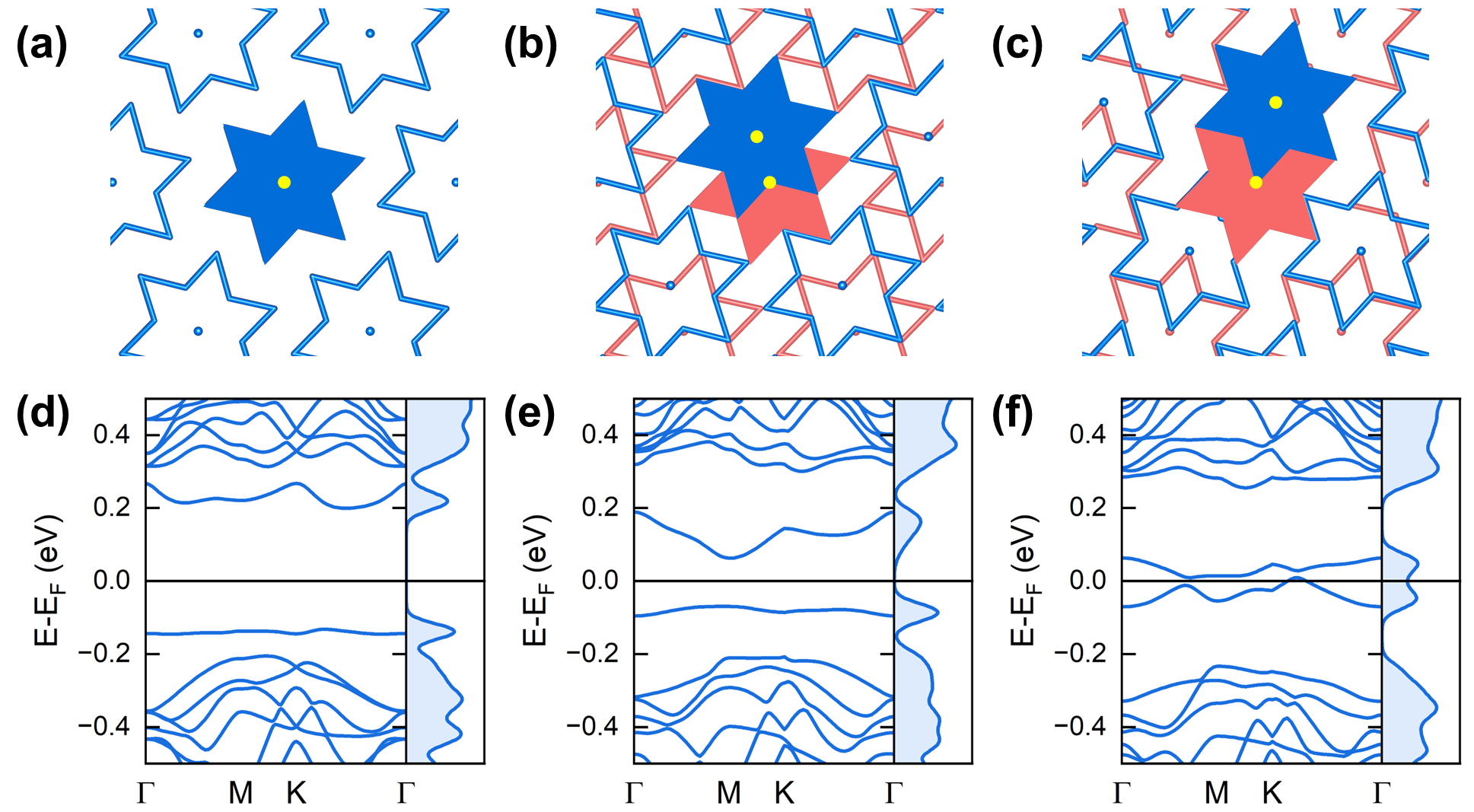}
    \caption{\textbf{The variation of the hybridization gap in bilayer T/T-TaS$_2$ due to interlayer stacking.}
    ({\bf{a-c}}) The top view of atomic structures for bilayer T/T-TaS$_2$ with T$_\textrm{A}$, T$_\textrm{B}$, and T$_\textrm{C}$ stacking, respectively. Two SoD clusters by Ta atoms in each layer are depicted in blue and red stars. All S atoms are omitted for clarity. ({\bf{d-f}}) Corresponding electronic band structures. All bands are doubly degenerate due to the inversion and time-reversal symmetries. From T$_\textrm{A}$, to T$_\textrm{B}$, and T$_\textrm{C}$, the bandwidth increases due to enhanced in-plane hopping and the band gap reduces.
    }
    \label{fig:fig2-bilayerbndstr}
\end{figure}

\textit{Results and Discussion} --
1T-TaS$_2$ undergoes a CDW transition to a commensurate $\sqrt{13}\times\sqrt{13}$ superlattice around 200 K~\cite{wang2020band} where twelve Ta atoms surrounding a central Ta are displaced towards the center. This results in a SoD structure that opens a gap of approximately 0.5 eV, which arises as the twelve Ta $5d^1$ electrons form a molecular orbital~\cite{darancet2014three,butler2020mottness}. The remaining spins localized at the center of the SoD form a triangular lattice in the 1T-TaS$_2$ superlattice and become Mott-localized~\cite{fazekas1979electrical,law20171t}. Our calculations demonstrate a band gap of about 0.3 eV within the CDW gap, as shown in Fig.~\ref{fig:fig1-str}(b). We note that we find the SoD structure to be stable in monolayers, T/T bilayers and T/T/H trilayers.

When two 1T layers are stacked, the number of electrons in the unit cell becomes even. Since the two layers are assembled with the two SoD structures facing each other, localized spins in opposite layers form a singlet state due to the interlayer hybridization. This results in a band insulator, as illustrated in Fig.~\ref{fig:fig1-str}(a,c)~\cite{darancet2014three,ritschel2018stacking} with fully filled bonding- and empty antibonding-bands. As expected, the band gap of the T/T bilayer is only marginally affected by on-site repulsion $U$, whereas 1T monolayer, being a genuine Mott insulator (half-filled band), exhibits a clear dependence on $U$ (as discussed in Supporting Information 1). We note that the top valence band (bonding-band) is even flatter in the T/T bilayer than in the 1T monolayer and, upon hole doping, correlation effects are expected to be significant, turning it into a (doped) Mott band.

Next we show that the T/T bilayer band structure is extremely sensitive to the interlayer stacking order. In a 1T monolayer, thirteen Ta atoms in a SoD are classified into three groups~\cite{ritschel2018stacking,lee2019origin}: one central atom, inner hexagon atoms (six atoms), and outer corners atoms (six atoms). When two 1T layers are stacked, the Ta atoms can be categorized into three groups based on the in-plane shift between two SoD. We note the direct stacking with no shift $\mathbf{T}_\textrm{A}$, shift to the inner hexagon $\mathbf{T}_\textrm{B}$ and shift to the outer corner $\mathbf{T}_\textrm{C}$, as illustrated in Fig.~\ref{fig:fig2-bilayerbndstr}(a-c). The lateral displacement reduces the inter-SoD coupling~\cite{butler2020mottness} but enhances the in-plane hopping~\cite{ritschel2015orbital,ritschel2018stacking,pizarro2020deconfinement}. Consequently, from T$_\textrm{A}$ to T$_\textrm{B}$ and T$_\textrm{C}$, the top valence and bottom conduction bands increase in bandwidth and their energy gap decreases. The $\mathbf{T}_\textrm{C}$ case even shows a zero indirect gap (see Fig.~\ref{fig:fig2-bilayerbndstr}(f)). Invoking the concept of chemical hardness, which stipulates the resistance to electron transfer in a system is proportional to the band gap~\cite{parr1983absolute}, we expect that the three stacking orders will exhibit distinct charge transfer from the T/T bilayer to the 1H layer upon the formation of the T/T/H heterostructure.
In the T/T bilayer, the T$_\textrm{A}$ stacking is the most stable in comparison to the T$_\textrm{B}$ and T$_\textrm{C}$ stackings, with relative formation energies of 83 and 159 meV per SoD pair, respectively. A previous DFT study on bulk 1T-TaS$_2$~\cite{lee2019origin} showed that charge doping or lattice strain can change the relative energy between T$_\textrm{A}$ and T$_\textrm{C}$ stackings. This indicates that the stacking energetics in the T/T/H trilayer may deviate from the T/T bilayer case.

\begin{figure}[t]
    \centering
    \includegraphics[width=1.0\columnwidth]{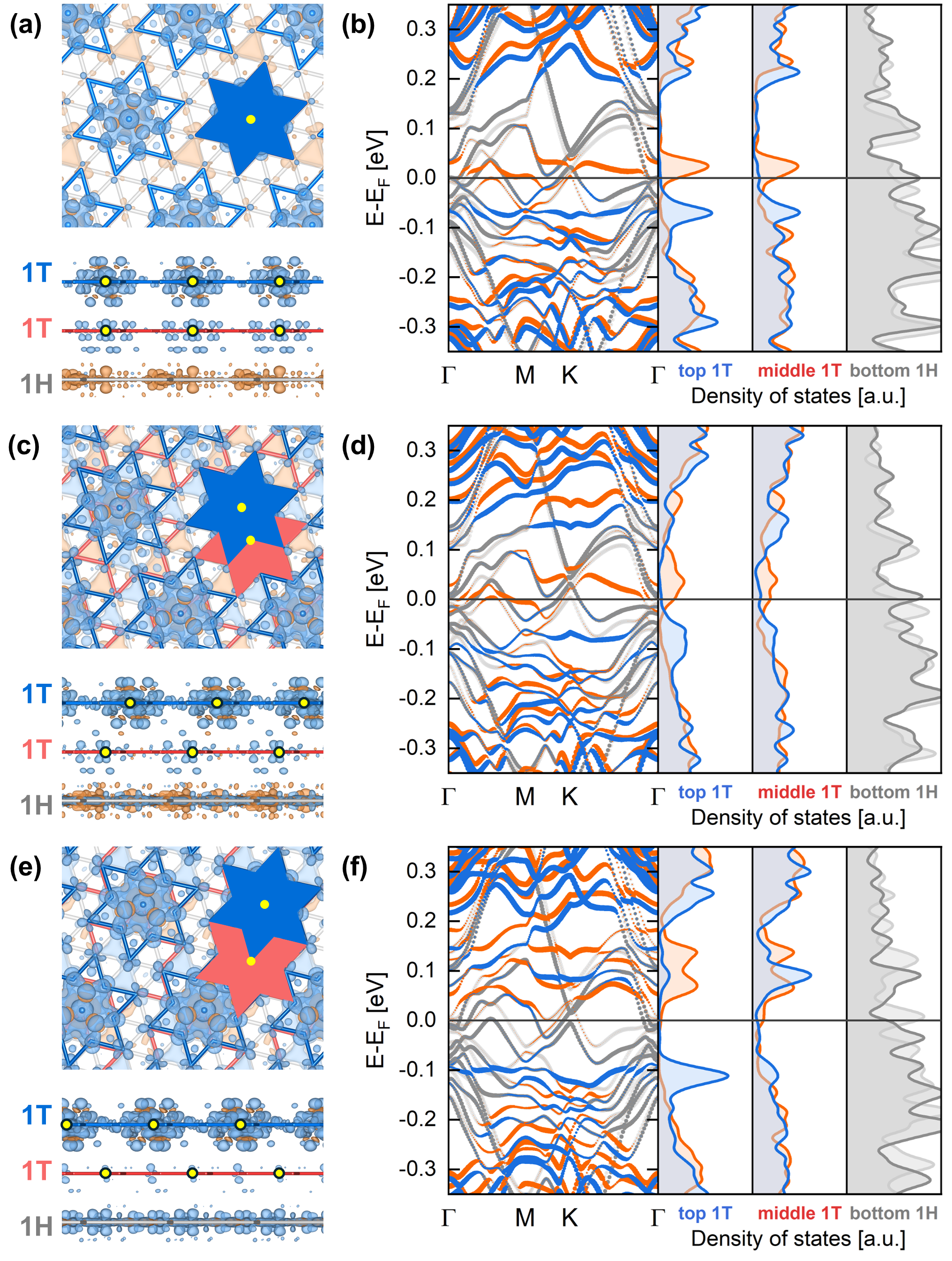}
    \caption{\textbf{Spin density and band structures of trilayer T/T/H-TaS$_2$ with different T/T stacking orders.}
    ({\bf{a}}) Top view (upper panel) and side view (lower panel) of spin density for the T$_\textrm{A}$ stacking. Blue and orange regions indicate opposite spin orientations. SoD clusters in the top and middle 1T layers are outlined in blue and red lines, and the triangular lattice in 1H layer is shown in gray lines. The central Ta atom is highlighed in yellow. ({\bf{b}}) Band structure and layer-resolved density of states for the T$_\textrm{A}$ trilayer. Majority and minority spin states on two 1T layers are depicted in blue and orange lines, while spin states in 1H layer are represented by dark and dim gray lines. ({\bf{c,d}}) Corresponding plots for the T$_\textrm{B}$, and ({\bf{e,f}}) for the T$_\textrm{C}$ stacking.
    }
    \label{fig:fig3-trilayerbndstr}
\end{figure}

We next investigate the electronic behavior of a trilayer heterostructure formed by a T/T bilayer in proximity to a 1H monolayer. The supercell is constrained to a $\sqrt{13}\times\sqrt{13}$ superlattice to incorporate the SoD structure. The $3\times3$ CDW in the 1H layer, which exhibits only marginal differences in charge transfer compared to $1\times1$ 1H~\cite{crippa2024heavy}, is neglected in these calculations. The interlayer separations in the optimal structures between the bottom 1H layer and the middle 1T layer is about $5.82\pm{}0.01$~\AA{}, close to that of T/H bilayer, of $5.81$~\AA{}~\cite{crippa2024heavy}. And the optimal distances between the 1T layers in the T/T/H trilayer and in the T/T bilayer are the same ($5.85\pm{}0.01$~\AA{}) for the three kinds of stacking orders. The resulting charge redistribution in the T/T/H trilayer can be rationalized by the competition between the interlayer interactions of the middle T layer with the top T layer and the bottom H layer. The electronic structures of the three trilayer heterostructures considered are shown in Fig.~\ref{fig:fig3-trilayerbndstr}. The T$_\textrm{C}$ case (Fig.~\ref{fig:fig3-trilayerbndstr}(e,f)) can be straightforwardly explained. Because of weak T/T layer coupling, the middle T layer loses one electron to the H layer and therefore the top T layer keeps approximately one unpaired electron per SoD. Consequently, the top T layer recovers a flat band Mott state, and the H layer exhibits weak spin polarization near the Fermi surface (Fig.~\ref{fig:fig3-trilayerbndstr}(f)). Due to reduced hopping in the plane, the new Mott band is flatter than for the corresponding T$_\textrm{C}$ stacking in the T/T bilayer case. Although the middle T layer becomes nearly a band insulator, it mediates the top-bottom coupling for the coexistence of localized spins in Mott bands and itinerant electrons in metallic bands, suggesting a possibility of heavy-fermion behavior. In the presence of superconductivity, the induced spin polarization in the 1H layer may favor $p$-wave Cooper pairing~\cite{Read2000,Ivanov2001,Duckheim2011,Chung2011}.

The T$_\textrm{A}$ stacking case (Fig.~\ref{fig:fig3-trilayerbndstr}(a,b)) exhibits slightly smaller electron transfer of 0.8 electron to the H layer than the T$_\textrm{C}$ stacking because of the stronger T/T hybridization in T$_\textrm{A}$ (details in Supporting Information 2). The original T/T bilayer valence band is slightly more than half filled, that is, the lower Hubbard band is fully occupied while the upper one is marginally occupied as shown in Fig.~\ref{fig:fig3-trilayerbndstr}(b). Despite the two T layers exhibiting similar density of states, the spin density in Fig.~\ref{fig:fig3-trilayerbndstr}(a) indicates that the middle T layer has a lower spin density than the top T layer due to the proximity of the H layer. The T$_\textrm{A}$ trilayer results therefore into an electron-doped Mott insulator, similar to the case of T/H bilayer which is a hole-doped Mott insulator~\cite{crippa2024heavy}. 
The T$_\textrm{B}$ stacking case shows a band structure and spin density intermediate to the T$_\textrm{A}$ and T$_\textrm{C}$ cases, as expected from the intermediate T/T coupling. Different from a T$_\textrm{C}$ stacking, T$_\textrm{A}$ and T$_\textrm{B}$ bilayer stackings induce opposite spin densities on the H layer compared to the spin density in T layers. Overall, T$_\textrm{A}$, T$_\textrm{B}$ and T$_\textrm{C}$ trilayers may be candidates for a continuous transition from a doped Mott insulator to a Kondo insulator behavior. 

The relative energetic stability between different T/T stacking indeed changes when forming the T/T/H trilayer. Compared to the bilayer case where the T$_\textrm{A}$ stacking is the most stable, the T$_\textrm{C}$ stacking is the energetically most favorable in the trilayer while T$_\textrm{A}$ and T$_\textrm{B}$ stackings are, respectively, 28 meV and 33 meV per SoD pair higher in energy. The close formation energies indicate that the three stacking orders may coexist in experiments, for example, in thermal annealing samples~\cite{wang2018surface}.

Beyond the trilayer T/T/H-TaS$_2$ heterostructure, bulky heterostructures composed of 1H layers and multilayer 1T layers thicker than a T/T bilayer are at least achievable through thermal processing~\cite{wang2018surface,sung2022two,sung2024endotaxial,husremovic2023encoding} and laser exposure~\cite{ravnik2019strain,ravnik2021quantum}, and are expected to host localized spin moments, unlike 4H$_\textrm{b}$-TaS$_2$ which has a well-defined alternative stacking order of 1H and 1T monolayers. Furthermore, if a fourth 1T layer is placed on the top of a T/T/H trilayer in a T$_\textrm{C}$ stacking, it might lead to a band insulator when the fourth and third T layers form a new band insulator bilayer. We can expect a rich phase diagram regarding different number of layers and varied interlayer stacking.

\textit{Conclusions} --
In conclusion, we demonstrated that TaS$_2$ trilayers combining a T/T bilayer and 1H monolayer exhibit flat bands with localized spins and itinerant states. The amount of charge transfer between the T/T bilayer and the 1H monolayer and resulting band structures depend on the T/T stacking order. Coexisting with itinerant electrons from the H layer, the T$_\textrm{C}$-type stacking results in a half-filled Mott band while the T$_\textrm{A,B}$-type stacking induces a partially doped Mott state. In contrast to the T/H bilayer, the 1H layer in the trilayer exhibits weak spin polarization following the charge transfer. These trilayer heterostructures facilitate a rich platform to study strong electron correlation and exotic superconductivity in vdW heterostructures.

\textit{Acknowledgments} --
B.Y. acknowledged the financial support by the Israel Science Foundation (ISF: 2932/21, 2974/23), German Research Foundation (DFG, CRC-183, A02), and by a research grant from the Estate of Gerald Alexander. I.I.M. was supported by the Office of Naval Research through the grant N00014-23-1-2480. R.V. acknowledges support by the Deutsche Forschungsgemeinschaft (DFG) through QUAST-FOR5249 - 449872909 (Project TP4) and Project No. VA117/23-1 — 509751747.

%\bibliography{reference}

\begin{thebibliography}{57}%
\makeatletter
\providecommand \@ifxundefined [1]{%
 \@ifx{#1\undefined}
}%
\providecommand \@ifnum [1]{%
 \ifnum #1\expandafter \@firstoftwo
 \else \expandafter \@secondoftwo
 \fi
}%
\providecommand \@ifx [1]{%
 \ifx #1\expandafter \@firstoftwo
 \else \expandafter \@secondoftwo
 \fi
}%
\providecommand \natexlab [1]{#1}%
\providecommand \enquote  [1]{``#1''}%
\providecommand \bibnamefont  [1]{#1}%
\providecommand \bibfnamefont [1]{#1}%
\providecommand \citenamefont [1]{#1}%
\providecommand \href@noop [0]{\@secondoftwo}%
\providecommand \href [0]{\begingroup \@sanitize@url \@href}%
\providecommand \@href[1]{\@@startlink{#1}\@@href}%
\providecommand \@@href[1]{\endgroup#1\@@endlink}%
\providecommand \@sanitize@url [0]{\catcode `\\12\catcode `\$12\catcode
  `\&12\catcode `\#12\catcode `\^12\catcode `\_12\catcode `\%12\relax}%
\providecommand \@@startlink[1]{}%
\providecommand \@@endlink[0]{}%
\providecommand \url  [0]{\begingroup\@sanitize@url \@url }%
\providecommand \@url [1]{\endgroup\@href {#1}{\urlprefix }}%
\providecommand \urlprefix  [0]{URL }%
\providecommand \Eprint [0]{\href }%
\providecommand \doibase [0]{https://doi.org/}%
\providecommand \selectlanguage [0]{\@gobble}%
\providecommand \bibinfo  [0]{\@secondoftwo}%
\providecommand \bibfield  [0]{\@secondoftwo}%
\providecommand \translation [1]{[#1]}%
\providecommand \BibitemOpen [0]{}%
\providecommand \bibitemStop [0]{}%
\providecommand \bibitemNoStop [0]{.\EOS\space}%
\providecommand \EOS [0]{\spacefactor3000\relax}%
\providecommand \BibitemShut  [1]{\csname bibitem#1\endcsname}%
\let\auto@bib@innerbib\@empty
%</preamble>
\bibitem [{\citenamefont {Geim}\ and\ \citenamefont
  {Grigorieva}(2013)}]{geim2013van}%
  \BibitemOpen
  \bibfield  {author} {\bibinfo {author} {\bibfnamefont {A.~K.}\ \bibnamefont
  {Geim}}\ and\ \bibinfo {author} {\bibfnamefont {I.~V.}\ \bibnamefont
  {Grigorieva}},\ }\href@noop {} {\bibfield  {journal} {\bibinfo  {journal}
  {Nature}\ }\textbf {\bibinfo {volume} {499}},\ \bibinfo {pages} {419}
  (\bibinfo {year} {2013})}\BibitemShut {NoStop}%
\bibitem [{\citenamefont {Novoselov}\ \emph {et~al.}(2016)\citenamefont
  {Novoselov}, \citenamefont {Mishchenko}, \citenamefont {Carvalho},\ and\
  \citenamefont {Castro~Neto}}]{novoselov20162d}%
  \BibitemOpen
  \bibfield  {author} {\bibinfo {author} {\bibfnamefont {K.~S.}\ \bibnamefont
  {Novoselov}}, \bibinfo {author} {\bibfnamefont {A.}~\bibnamefont
  {Mishchenko}}, \bibinfo {author} {\bibfnamefont {A.}~\bibnamefont
  {Carvalho}},\ and\ \bibinfo {author} {\bibfnamefont {A.}~\bibnamefont
  {Castro~Neto}},\ }\href@noop {} {\bibfield  {journal} {\bibinfo  {journal}
  {Science}\ }\textbf {\bibinfo {volume} {353}},\ \bibinfo {pages} {aac9439}
  (\bibinfo {year} {2016})}\BibitemShut {NoStop}%
\bibitem [{\citenamefont {Balents}\ \emph {et~al.}(2020)\citenamefont
  {Balents}, \citenamefont {Dean}, \citenamefont {Efetov},\ and\ \citenamefont
  {Young}}]{balents2020superconductivity}%
  \BibitemOpen
  \bibfield  {author} {\bibinfo {author} {\bibfnamefont {L.}~\bibnamefont
  {Balents}}, \bibinfo {author} {\bibfnamefont {C.~R.}\ \bibnamefont {Dean}},
  \bibinfo {author} {\bibfnamefont {D.~K.}\ \bibnamefont {Efetov}},\ and\
  \bibinfo {author} {\bibfnamefont {A.~F.}\ \bibnamefont {Young}},\ }\href@noop
  {} {\bibfield  {journal} {\bibinfo  {journal} {Nature Physics}\ }\textbf
  {\bibinfo {volume} {16}},\ \bibinfo {pages} {725} (\bibinfo {year}
  {2020})}\BibitemShut {NoStop}%
\bibitem [{\citenamefont {Bistritzer}\ and\ \citenamefont
  {MacDonald}(2011)}]{bistritzer2011moire}%
  \BibitemOpen
  \bibfield  {author} {\bibinfo {author} {\bibfnamefont {R.}~\bibnamefont
  {Bistritzer}}\ and\ \bibinfo {author} {\bibfnamefont {A.~H.}\ \bibnamefont
  {MacDonald}},\ }\href@noop {} {\bibfield  {journal} {\bibinfo  {journal}
  {Proceedings of the National Academy of Sciences}\ }\textbf {\bibinfo
  {volume} {108}},\ \bibinfo {pages} {12233} (\bibinfo {year}
  {2011})}\BibitemShut {NoStop}%
\bibitem [{\citenamefont {Cao}\ \emph {et~al.}(2018)\citenamefont {Cao},
  \citenamefont {Fatemi}, \citenamefont {Fang}, \citenamefont {Watanabe},
  \citenamefont {Taniguchi}, \citenamefont {Kaxiras},\ and\ \citenamefont
  {Jarillo-Herrero}}]{cao2018unconventional}%
  \BibitemOpen
  \bibfield  {author} {\bibinfo {author} {\bibfnamefont {Y.}~\bibnamefont
  {Cao}}, \bibinfo {author} {\bibfnamefont {V.}~\bibnamefont {Fatemi}},
  \bibinfo {author} {\bibfnamefont {S.}~\bibnamefont {Fang}}, \bibinfo {author}
  {\bibfnamefont {K.}~\bibnamefont {Watanabe}}, \bibinfo {author}
  {\bibfnamefont {T.}~\bibnamefont {Taniguchi}}, \bibinfo {author}
  {\bibfnamefont {E.}~\bibnamefont {Kaxiras}},\ and\ \bibinfo {author}
  {\bibfnamefont {P.}~\bibnamefont {Jarillo-Herrero}},\ }\href@noop {}
  {\bibfield  {journal} {\bibinfo  {journal} {Nature}\ }\textbf {\bibinfo
  {volume} {556}},\ \bibinfo {pages} {43} (\bibinfo {year} {2018})}\BibitemShut
  {NoStop}%
\bibitem [{\citenamefont {Jiang}\ \emph {et~al.}(2019)\citenamefont {Jiang},
  \citenamefont {Lai}, \citenamefont {Watanabe}, \citenamefont {Taniguchi},
  \citenamefont {Haule}, \citenamefont {Mao},\ and\ \citenamefont
  {Andrei}}]{jiang2019charge}%
  \BibitemOpen
  \bibfield  {author} {\bibinfo {author} {\bibfnamefont {Y.}~\bibnamefont
  {Jiang}}, \bibinfo {author} {\bibfnamefont {X.}~\bibnamefont {Lai}}, \bibinfo
  {author} {\bibfnamefont {K.}~\bibnamefont {Watanabe}}, \bibinfo {author}
  {\bibfnamefont {T.}~\bibnamefont {Taniguchi}}, \bibinfo {author}
  {\bibfnamefont {K.}~\bibnamefont {Haule}}, \bibinfo {author} {\bibfnamefont
  {J.}~\bibnamefont {Mao}},\ and\ \bibinfo {author} {\bibfnamefont {E.~Y.}\
  \bibnamefont {Andrei}},\ }\href@noop {} {\bibfield  {journal} {\bibinfo
  {journal} {Nature}\ }\textbf {\bibinfo {volume} {573}},\ \bibinfo {pages}
  {91} (\bibinfo {year} {2019})}\BibitemShut {NoStop}%
\bibitem [{\citenamefont {Wilson}\ \emph {et~al.}(2001)\citenamefont {Wilson},
  \citenamefont {Di~Salvo},\ and\ \citenamefont {Mahajan}}]{wilson2001charge}%
  \BibitemOpen
  \bibfield  {author} {\bibinfo {author} {\bibfnamefont {J.}~\bibnamefont
  {Wilson}}, \bibinfo {author} {\bibfnamefont {F.}~\bibnamefont {Di~Salvo}},\
  and\ \bibinfo {author} {\bibfnamefont {S.}~\bibnamefont {Mahajan}},\
  }\href@noop {} {\bibfield  {journal} {\bibinfo  {journal} {Advances in
  Physics}\ }\textbf {\bibinfo {volume} {50}},\ \bibinfo {pages} {1171}
  (\bibinfo {year} {2001})}\BibitemShut {NoStop}%
\bibitem [{\citenamefont {Yang}\ \emph {et~al.}(2017)\citenamefont {Yang},
  \citenamefont {Kim}, \citenamefont {Chhowalla},\ and\ \citenamefont
  {Lee}}]{yang2017structural}%
  \BibitemOpen
  \bibfield  {author} {\bibinfo {author} {\bibfnamefont {H.}~\bibnamefont
  {Yang}}, \bibinfo {author} {\bibfnamefont {S.~W.}\ \bibnamefont {Kim}},
  \bibinfo {author} {\bibfnamefont {M.}~\bibnamefont {Chhowalla}},\ and\
  \bibinfo {author} {\bibfnamefont {Y.~H.}\ \bibnamefont {Lee}},\ }\href@noop
  {} {\bibfield  {journal} {\bibinfo  {journal} {Nature Physics}\ }\textbf
  {\bibinfo {volume} {13}},\ \bibinfo {pages} {931} (\bibinfo {year}
  {2017})}\BibitemShut {NoStop}%
\bibitem [{\citenamefont {Wang}\ \emph {et~al.}(2020)\citenamefont {Wang},
  \citenamefont {Yao}, \citenamefont {Xin}, \citenamefont {Han}, \citenamefont
  {Wang}, \citenamefont {Chen}, \citenamefont {Cai}, \citenamefont {Li},\ and\
  \citenamefont {Zhang}}]{wang2020band}%
  \BibitemOpen
  \bibfield  {author} {\bibinfo {author} {\bibfnamefont {Y.}~\bibnamefont
  {Wang}}, \bibinfo {author} {\bibfnamefont {W.}~\bibnamefont {Yao}}, \bibinfo
  {author} {\bibfnamefont {Z.}~\bibnamefont {Xin}}, \bibinfo {author}
  {\bibfnamefont {T.}~\bibnamefont {Han}}, \bibinfo {author} {\bibfnamefont
  {Z.}~\bibnamefont {Wang}}, \bibinfo {author} {\bibfnamefont {L.}~\bibnamefont
  {Chen}}, \bibinfo {author} {\bibfnamefont {C.}~\bibnamefont {Cai}}, \bibinfo
  {author} {\bibfnamefont {Y.}~\bibnamefont {Li}},\ and\ \bibinfo {author}
  {\bibfnamefont {Y.}~\bibnamefont {Zhang}},\ }\href@noop {} {\bibfield
  {journal} {\bibinfo  {journal} {Nature communications}\ }\textbf {\bibinfo
  {volume} {11}},\ \bibinfo {pages} {4215} (\bibinfo {year}
  {2020})}\BibitemShut {NoStop}%
\bibitem [{\citenamefont {Chen}\ \emph {et~al.}(2020)\citenamefont {Chen},
  \citenamefont {Ruan}, \citenamefont {Wu}, \citenamefont {Tang}, \citenamefont
  {Ryu}, \citenamefont {Tsai}, \citenamefont {Lee}, \citenamefont {Kahn},
  \citenamefont {Liou}, \citenamefont {Jia} \emph {et~al.}}]{chen2020strong}%
  \BibitemOpen
  \bibfield  {author} {\bibinfo {author} {\bibfnamefont {Y.}~\bibnamefont
  {Chen}}, \bibinfo {author} {\bibfnamefont {W.}~\bibnamefont {Ruan}}, \bibinfo
  {author} {\bibfnamefont {M.}~\bibnamefont {Wu}}, \bibinfo {author}
  {\bibfnamefont {S.}~\bibnamefont {Tang}}, \bibinfo {author} {\bibfnamefont
  {H.}~\bibnamefont {Ryu}}, \bibinfo {author} {\bibfnamefont {H.-Z.}\
  \bibnamefont {Tsai}}, \bibinfo {author} {\bibfnamefont {R.~L.}\ \bibnamefont
  {Lee}}, \bibinfo {author} {\bibfnamefont {S.}~\bibnamefont {Kahn}}, \bibinfo
  {author} {\bibfnamefont {F.}~\bibnamefont {Liou}}, \bibinfo {author}
  {\bibfnamefont {C.}~\bibnamefont {Jia}}, \emph {et~al.},\ }\href@noop {}
  {\bibfield  {journal} {\bibinfo  {journal} {Nature Physics}\ }\textbf
  {\bibinfo {volume} {16}},\ \bibinfo {pages} {218} (\bibinfo {year}
  {2020})}\BibitemShut {NoStop}%
\bibitem [{\citenamefont {Nakata}\ \emph {et~al.}(2021)\citenamefont {Nakata},
  \citenamefont {Sugawara}, \citenamefont {Chainani}, \citenamefont {Oka},
  \citenamefont {Bao}, \citenamefont {Zhou}, \citenamefont {Chuang},
  \citenamefont {Cheng}, \citenamefont {Kawakami}, \citenamefont {Saruta} \emph
  {et~al.}}]{nakata2021robust}%
  \BibitemOpen
  \bibfield  {author} {\bibinfo {author} {\bibfnamefont {Y.}~\bibnamefont
  {Nakata}}, \bibinfo {author} {\bibfnamefont {K.}~\bibnamefont {Sugawara}},
  \bibinfo {author} {\bibfnamefont {A.}~\bibnamefont {Chainani}}, \bibinfo
  {author} {\bibfnamefont {H.}~\bibnamefont {Oka}}, \bibinfo {author}
  {\bibfnamefont {C.}~\bibnamefont {Bao}}, \bibinfo {author} {\bibfnamefont
  {S.}~\bibnamefont {Zhou}}, \bibinfo {author} {\bibfnamefont {P.-Y.}\
  \bibnamefont {Chuang}}, \bibinfo {author} {\bibfnamefont {C.-M.}\
  \bibnamefont {Cheng}}, \bibinfo {author} {\bibfnamefont {T.}~\bibnamefont
  {Kawakami}}, \bibinfo {author} {\bibfnamefont {Y.}~\bibnamefont {Saruta}},
  \emph {et~al.},\ }\href@noop {} {\bibfield  {journal} {\bibinfo  {journal}
  {Nature communications}\ }\textbf {\bibinfo {volume} {12}},\ \bibinfo {pages}
  {5873} (\bibinfo {year} {2021})}\BibitemShut {NoStop}%
\bibitem [{\citenamefont {Law}\ and\ \citenamefont {Lee}(2017)}]{law20171t}%
  \BibitemOpen
  \bibfield  {author} {\bibinfo {author} {\bibfnamefont {K.~T.}\ \bibnamefont
  {Law}}\ and\ \bibinfo {author} {\bibfnamefont {P.~A.}\ \bibnamefont {Lee}},\
  }\href@noop {} {\bibfield  {journal} {\bibinfo  {journal} {Proceedings of the
  National Academy of Sciences}\ }\textbf {\bibinfo {volume} {114}},\ \bibinfo
  {pages} {6996} (\bibinfo {year} {2017})}\BibitemShut {NoStop}%
\bibitem [{\citenamefont {He}\ \emph {et~al.}(2018)\citenamefont {He},
  \citenamefont {Xu}, \citenamefont {Chen}, \citenamefont {Law},\ and\
  \citenamefont {Lee}}]{he2018spinon}%
  \BibitemOpen
  \bibfield  {author} {\bibinfo {author} {\bibfnamefont {W.-Y.}\ \bibnamefont
  {He}}, \bibinfo {author} {\bibfnamefont {X.~Y.}\ \bibnamefont {Xu}}, \bibinfo
  {author} {\bibfnamefont {G.}~\bibnamefont {Chen}}, \bibinfo {author}
  {\bibfnamefont {K.~T.}\ \bibnamefont {Law}},\ and\ \bibinfo {author}
  {\bibfnamefont {P.~A.}\ \bibnamefont {Lee}},\ }\href@noop {} {\bibfield
  {journal} {\bibinfo  {journal} {Physical review letters}\ }\textbf {\bibinfo
  {volume} {121}},\ \bibinfo {pages} {046401} (\bibinfo {year}
  {2018})}\BibitemShut {NoStop}%
\bibitem [{\citenamefont {Ribak}\ \emph {et~al.}(2017)\citenamefont {Ribak},
  \citenamefont {Silber}, \citenamefont {Baines}, \citenamefont {Chashka},
  \citenamefont {Salman}, \citenamefont {Dagan},\ and\ \citenamefont
  {Kanigel}}]{ribak2017gapless}%
  \BibitemOpen
  \bibfield  {author} {\bibinfo {author} {\bibfnamefont {A.}~\bibnamefont
  {Ribak}}, \bibinfo {author} {\bibfnamefont {I.}~\bibnamefont {Silber}},
  \bibinfo {author} {\bibfnamefont {C.}~\bibnamefont {Baines}}, \bibinfo
  {author} {\bibfnamefont {K.}~\bibnamefont {Chashka}}, \bibinfo {author}
  {\bibfnamefont {Z.}~\bibnamefont {Salman}}, \bibinfo {author} {\bibfnamefont
  {Y.}~\bibnamefont {Dagan}},\ and\ \bibinfo {author} {\bibfnamefont
  {A.}~\bibnamefont {Kanigel}},\ }\href@noop {} {\bibfield  {journal} {\bibinfo
   {journal} {Physical Review B}\ }\textbf {\bibinfo {volume} {96}},\ \bibinfo
  {pages} {195131} (\bibinfo {year} {2017})}\BibitemShut {NoStop}%
\bibitem [{\citenamefont {Klanj{\v{s}}ek}\ \emph {et~al.}(2017)\citenamefont
  {Klanj{\v{s}}ek}, \citenamefont {Zorko}, \citenamefont {{\v{Z}}itko},
  \citenamefont {Mravlje}, \citenamefont {Jagli{\v{c}}i{\'c}}, \citenamefont
  {Biswas}, \citenamefont {Prelov{\v{s}}ek}, \citenamefont {Mihailovic},\ and\
  \citenamefont {Ar{\v{c}}on}}]{klanjvsek2017high}%
  \BibitemOpen
  \bibfield  {author} {\bibinfo {author} {\bibfnamefont {M.}~\bibnamefont
  {Klanj{\v{s}}ek}}, \bibinfo {author} {\bibfnamefont {A.}~\bibnamefont
  {Zorko}}, \bibinfo {author} {\bibfnamefont {R.}~\bibnamefont {{\v{Z}}itko}},
  \bibinfo {author} {\bibfnamefont {J.}~\bibnamefont {Mravlje}}, \bibinfo
  {author} {\bibfnamefont {Z.}~\bibnamefont {Jagli{\v{c}}i{\'c}}}, \bibinfo
  {author} {\bibfnamefont {P.~K.}\ \bibnamefont {Biswas}}, \bibinfo {author}
  {\bibfnamefont {P.}~\bibnamefont {Prelov{\v{s}}ek}}, \bibinfo {author}
  {\bibfnamefont {D.}~\bibnamefont {Mihailovic}},\ and\ \bibinfo {author}
  {\bibfnamefont {D.}~\bibnamefont {Ar{\v{c}}on}},\ }\href@noop {} {\bibfield
  {journal} {\bibinfo  {journal} {Nature Physics}\ }\textbf {\bibinfo {volume}
  {13}},\ \bibinfo {pages} {1130} (\bibinfo {year} {2017})}\BibitemShut
  {NoStop}%
\bibitem [{\citenamefont {Murayama}\ \emph {et~al.}(2020)\citenamefont
  {Murayama}, \citenamefont {Sato}, \citenamefont {Taniguchi}, \citenamefont
  {Kurihara}, \citenamefont {Xing}, \citenamefont {Huang}, \citenamefont
  {Kasahara}, \citenamefont {Kasahara}, \citenamefont {Kimchi}, \citenamefont
  {Yoshida} \emph {et~al.}}]{murayama2020effect}%
  \BibitemOpen
  \bibfield  {author} {\bibinfo {author} {\bibfnamefont {H.}~\bibnamefont
  {Murayama}}, \bibinfo {author} {\bibfnamefont {Y.}~\bibnamefont {Sato}},
  \bibinfo {author} {\bibfnamefont {T.}~\bibnamefont {Taniguchi}}, \bibinfo
  {author} {\bibfnamefont {R.}~\bibnamefont {Kurihara}}, \bibinfo {author}
  {\bibfnamefont {X.}~\bibnamefont {Xing}}, \bibinfo {author} {\bibfnamefont
  {W.}~\bibnamefont {Huang}}, \bibinfo {author} {\bibfnamefont
  {S.}~\bibnamefont {Kasahara}}, \bibinfo {author} {\bibfnamefont
  {Y.}~\bibnamefont {Kasahara}}, \bibinfo {author} {\bibfnamefont
  {I.}~\bibnamefont {Kimchi}}, \bibinfo {author} {\bibfnamefont
  {M.}~\bibnamefont {Yoshida}}, \emph {et~al.},\ }\href@noop {} {\bibfield
  {journal} {\bibinfo  {journal} {Physical Review Research}\ }\textbf {\bibinfo
  {volume} {2}},\ \bibinfo {pages} {013099} (\bibinfo {year}
  {2020})}\BibitemShut {NoStop}%
\bibitem [{\citenamefont {Ma{\~n}as-Valero}\ \emph {et~al.}(2021)\citenamefont
  {Ma{\~n}as-Valero}, \citenamefont {Huddart}, \citenamefont {Lancaster},
  \citenamefont {Coronado},\ and\ \citenamefont {Pratt}}]{manas2021quantum}%
  \BibitemOpen
  \bibfield  {author} {\bibinfo {author} {\bibfnamefont {S.}~\bibnamefont
  {Ma{\~n}as-Valero}}, \bibinfo {author} {\bibfnamefont {B.~M.}\ \bibnamefont
  {Huddart}}, \bibinfo {author} {\bibfnamefont {T.}~\bibnamefont {Lancaster}},
  \bibinfo {author} {\bibfnamefont {E.}~\bibnamefont {Coronado}},\ and\
  \bibinfo {author} {\bibfnamefont {F.~L.}\ \bibnamefont {Pratt}},\ }\href@noop
  {} {\bibfield  {journal} {\bibinfo  {journal} {npj Quantum Materials}\
  }\textbf {\bibinfo {volume} {6}},\ \bibinfo {pages} {69} (\bibinfo {year}
  {2021})}\BibitemShut {NoStop}%
\bibitem [{\citenamefont {Benedi{\v{c}}i{\v{c}}}\ \emph
  {et~al.}(2020)\citenamefont {Benedi{\v{c}}i{\v{c}}}, \citenamefont
  {Jan{\v{s}}a}, \citenamefont {van Midden}, \citenamefont {Jegli{\v{c}}},
  \citenamefont {Klanj{\v{s}}ek}, \citenamefont {Zupani{\v{c}}}, \citenamefont
  {Jagli{\v{c}}i{\'c}}, \citenamefont {{\v{S}}utar}, \citenamefont
  {Prelov{\v{s}}ek}, \citenamefont {Mihailovi{\v{c}}} \emph
  {et~al.}}]{benedivcivc2020superconductivity}%
  \BibitemOpen
  \bibfield  {author} {\bibinfo {author} {\bibfnamefont {I.}~\bibnamefont
  {Benedi{\v{c}}i{\v{c}}}}, \bibinfo {author} {\bibfnamefont {N.}~\bibnamefont
  {Jan{\v{s}}a}}, \bibinfo {author} {\bibfnamefont {M.}~\bibnamefont {van
  Midden}}, \bibinfo {author} {\bibfnamefont {P.}~\bibnamefont {Jegli{\v{c}}}},
  \bibinfo {author} {\bibfnamefont {M.}~\bibnamefont {Klanj{\v{s}}ek}},
  \bibinfo {author} {\bibfnamefont {E.}~\bibnamefont {Zupani{\v{c}}}}, \bibinfo
  {author} {\bibfnamefont {Z.}~\bibnamefont {Jagli{\v{c}}i{\'c}}}, \bibinfo
  {author} {\bibfnamefont {P.}~\bibnamefont {{\v{S}}utar}}, \bibinfo {author}
  {\bibfnamefont {P.}~\bibnamefont {Prelov{\v{s}}ek}}, \bibinfo {author}
  {\bibfnamefont {D.}~\bibnamefont {Mihailovi{\v{c}}}}, \emph {et~al.},\
  }\href@noop {} {\bibfield  {journal} {\bibinfo  {journal} {Physical Review
  B}\ }\textbf {\bibinfo {volume} {102}},\ \bibinfo {pages} {054401} (\bibinfo
  {year} {2020})}\BibitemShut {NoStop}%
\bibitem [{\citenamefont {Ruan}\ \emph {et~al.}(2021)\citenamefont {Ruan},
  \citenamefont {Chen}, \citenamefont {Tang}, \citenamefont {Hwang},
  \citenamefont {Tsai}, \citenamefont {Lee}, \citenamefont {Wu}, \citenamefont
  {Ryu}, \citenamefont {Kahn}, \citenamefont {Liou} \emph
  {et~al.}}]{ruan2021evidence}%
  \BibitemOpen
  \bibfield  {author} {\bibinfo {author} {\bibfnamefont {W.}~\bibnamefont
  {Ruan}}, \bibinfo {author} {\bibfnamefont {Y.}~\bibnamefont {Chen}}, \bibinfo
  {author} {\bibfnamefont {S.}~\bibnamefont {Tang}}, \bibinfo {author}
  {\bibfnamefont {J.}~\bibnamefont {Hwang}}, \bibinfo {author} {\bibfnamefont
  {H.-Z.}\ \bibnamefont {Tsai}}, \bibinfo {author} {\bibfnamefont {R.~L.}\
  \bibnamefont {Lee}}, \bibinfo {author} {\bibfnamefont {M.}~\bibnamefont
  {Wu}}, \bibinfo {author} {\bibfnamefont {H.}~\bibnamefont {Ryu}}, \bibinfo
  {author} {\bibfnamefont {S.}~\bibnamefont {Kahn}}, \bibinfo {author}
  {\bibfnamefont {F.}~\bibnamefont {Liou}}, \emph {et~al.},\ }\href@noop {}
  {\bibfield  {journal} {\bibinfo  {journal} {Nature Physics}\ }\textbf
  {\bibinfo {volume} {17}},\ \bibinfo {pages} {1154} (\bibinfo {year}
  {2021})}\BibitemShut {NoStop}%
\bibitem [{\citenamefont {Chen}\ \emph {et~al.}(2022)\citenamefont {Chen},
  \citenamefont {He}, \citenamefont {Ruan}, \citenamefont {Hwang},
  \citenamefont {Tang}, \citenamefont {Lee}, \citenamefont {Wu}, \citenamefont
  {Zhu}, \citenamefont {Zhang}, \citenamefont {Ryu} \emph
  {et~al.}}]{chen2022evidence}%
  \BibitemOpen
  \bibfield  {author} {\bibinfo {author} {\bibfnamefont {Y.}~\bibnamefont
  {Chen}}, \bibinfo {author} {\bibfnamefont {W.-Y.}\ \bibnamefont {He}},
  \bibinfo {author} {\bibfnamefont {W.}~\bibnamefont {Ruan}}, \bibinfo {author}
  {\bibfnamefont {J.}~\bibnamefont {Hwang}}, \bibinfo {author} {\bibfnamefont
  {S.}~\bibnamefont {Tang}}, \bibinfo {author} {\bibfnamefont {R.~L.}\
  \bibnamefont {Lee}}, \bibinfo {author} {\bibfnamefont {M.}~\bibnamefont
  {Wu}}, \bibinfo {author} {\bibfnamefont {T.}~\bibnamefont {Zhu}}, \bibinfo
  {author} {\bibfnamefont {C.}~\bibnamefont {Zhang}}, \bibinfo {author}
  {\bibfnamefont {H.}~\bibnamefont {Ryu}}, \emph {et~al.},\ }\href@noop {}
  {\bibfield  {journal} {\bibinfo  {journal} {Nature Physics}\ }\textbf
  {\bibinfo {volume} {18}},\ \bibinfo {pages} {1335} (\bibinfo {year}
  {2022})}\BibitemShut {NoStop}%
\bibitem [{\citenamefont {Coleman}\ \emph {et~al.}(1983)\citenamefont
  {Coleman}, \citenamefont {Eiserman}, \citenamefont {Hillenius}, \citenamefont
  {Mitchell},\ and\ \citenamefont {Vicent}}]{coleman1983dimensional}%
  \BibitemOpen
  \bibfield  {author} {\bibinfo {author} {\bibfnamefont {R.}~\bibnamefont
  {Coleman}}, \bibinfo {author} {\bibfnamefont {G.}~\bibnamefont {Eiserman}},
  \bibinfo {author} {\bibfnamefont {S.}~\bibnamefont {Hillenius}}, \bibinfo
  {author} {\bibfnamefont {A.}~\bibnamefont {Mitchell}},\ and\ \bibinfo
  {author} {\bibfnamefont {J.}~\bibnamefont {Vicent}},\ }\href@noop {}
  {\bibfield  {journal} {\bibinfo  {journal} {Physical Review B}\ }\textbf
  {\bibinfo {volume} {27}},\ \bibinfo {pages} {125} (\bibinfo {year}
  {1983})}\BibitemShut {NoStop}%
\bibitem [{\citenamefont {De~la Barrera}\ \emph {et~al.}(2018)\citenamefont
  {De~la Barrera}, \citenamefont {Sinko}, \citenamefont {Gopalan},
  \citenamefont {Sivadas}, \citenamefont {Seyler}, \citenamefont {Watanabe},
  \citenamefont {Taniguchi}, \citenamefont {Tsen}, \citenamefont {Xu},
  \citenamefont {Xiao} \emph {et~al.}}]{de2018tuning}%
  \BibitemOpen
  \bibfield  {author} {\bibinfo {author} {\bibfnamefont {S.~C.}\ \bibnamefont
  {De~la Barrera}}, \bibinfo {author} {\bibfnamefont {M.~R.}\ \bibnamefont
  {Sinko}}, \bibinfo {author} {\bibfnamefont {D.~P.}\ \bibnamefont {Gopalan}},
  \bibinfo {author} {\bibfnamefont {N.}~\bibnamefont {Sivadas}}, \bibinfo
  {author} {\bibfnamefont {K.~L.}\ \bibnamefont {Seyler}}, \bibinfo {author}
  {\bibfnamefont {K.}~\bibnamefont {Watanabe}}, \bibinfo {author}
  {\bibfnamefont {T.}~\bibnamefont {Taniguchi}}, \bibinfo {author}
  {\bibfnamefont {A.~W.}\ \bibnamefont {Tsen}}, \bibinfo {author}
  {\bibfnamefont {X.}~\bibnamefont {Xu}}, \bibinfo {author} {\bibfnamefont
  {D.}~\bibnamefont {Xiao}}, \emph {et~al.},\ }\href@noop {} {\bibfield
  {journal} {\bibinfo  {journal} {Nature communications}\ }\textbf {\bibinfo
  {volume} {9}},\ \bibinfo {pages} {1427} (\bibinfo {year} {2018})}\BibitemShut
  {NoStop}%
\bibitem [{\citenamefont {Bhoi}\ \emph {et~al.}(2016)\citenamefont {Bhoi},
  \citenamefont {Khim}, \citenamefont {Nam}, \citenamefont {Lee}, \citenamefont
  {Kim}, \citenamefont {Jeon}, \citenamefont {Min}, \citenamefont {Park},\ and\
  \citenamefont {Kim}}]{bhoi2016interplay}%
  \BibitemOpen
  \bibfield  {author} {\bibinfo {author} {\bibfnamefont {D.}~\bibnamefont
  {Bhoi}}, \bibinfo {author} {\bibfnamefont {S.}~\bibnamefont {Khim}}, \bibinfo
  {author} {\bibfnamefont {W.}~\bibnamefont {Nam}}, \bibinfo {author}
  {\bibfnamefont {B.}~\bibnamefont {Lee}}, \bibinfo {author} {\bibfnamefont
  {C.}~\bibnamefont {Kim}}, \bibinfo {author} {\bibfnamefont {B.-G.}\
  \bibnamefont {Jeon}}, \bibinfo {author} {\bibfnamefont {B.}~\bibnamefont
  {Min}}, \bibinfo {author} {\bibfnamefont {S.}~\bibnamefont {Park}},\ and\
  \bibinfo {author} {\bibfnamefont {K.~H.}\ \bibnamefont {Kim}},\ }\href@noop
  {} {\bibfield  {journal} {\bibinfo  {journal} {Scientific reports}\ }\textbf
  {\bibinfo {volume} {6}},\ \bibinfo {pages} {24068} (\bibinfo {year}
  {2016})}\BibitemShut {NoStop}%
\bibitem [{\citenamefont {Lian}\ \emph {et~al.}(2019)\citenamefont {Lian},
  \citenamefont {Heil}, \citenamefont {Liu}, \citenamefont {Si}, \citenamefont
  {Giustino},\ and\ \citenamefont {Duan}}]{lian2019coexistence}%
  \BibitemOpen
  \bibfield  {author} {\bibinfo {author} {\bibfnamefont {C.-S.}\ \bibnamefont
  {Lian}}, \bibinfo {author} {\bibfnamefont {C.}~\bibnamefont {Heil}}, \bibinfo
  {author} {\bibfnamefont {X.}~\bibnamefont {Liu}}, \bibinfo {author}
  {\bibfnamefont {C.}~\bibnamefont {Si}}, \bibinfo {author} {\bibfnamefont
  {F.}~\bibnamefont {Giustino}},\ and\ \bibinfo {author} {\bibfnamefont
  {W.}~\bibnamefont {Duan}},\ }\href@noop {} {\bibfield  {journal} {\bibinfo
  {journal} {The Journal of Physical Chemistry Letters}\ }\textbf {\bibinfo
  {volume} {10}},\ \bibinfo {pages} {4076} (\bibinfo {year}
  {2019})}\BibitemShut {NoStop}%
\bibitem [{\citenamefont {Ribak}\ \emph {et~al.}(2020)\citenamefont {Ribak},
  \citenamefont {Skiff}, \citenamefont {Mograbi}, \citenamefont {Rout},
  \citenamefont {Fischer}, \citenamefont {Ruhman}, \citenamefont {Chashka},
  \citenamefont {Dagan},\ and\ \citenamefont {Kanigel}}]{ribak2020chiral}%
  \BibitemOpen
  \bibfield  {author} {\bibinfo {author} {\bibfnamefont {A.}~\bibnamefont
  {Ribak}}, \bibinfo {author} {\bibfnamefont {R.~M.}\ \bibnamefont {Skiff}},
  \bibinfo {author} {\bibfnamefont {M.}~\bibnamefont {Mograbi}}, \bibinfo
  {author} {\bibfnamefont {P.}~\bibnamefont {Rout}}, \bibinfo {author}
  {\bibfnamefont {M.}~\bibnamefont {Fischer}}, \bibinfo {author} {\bibfnamefont
  {J.}~\bibnamefont {Ruhman}}, \bibinfo {author} {\bibfnamefont
  {K.}~\bibnamefont {Chashka}}, \bibinfo {author} {\bibfnamefont
  {Y.}~\bibnamefont {Dagan}},\ and\ \bibinfo {author} {\bibfnamefont
  {A.}~\bibnamefont {Kanigel}},\ }\href@noop {} {\bibfield  {journal} {\bibinfo
   {journal} {Science advances}\ }\textbf {\bibinfo {volume} {6}},\ \bibinfo
  {pages} {eaax9480} (\bibinfo {year} {2020})}\BibitemShut {NoStop}%
\bibitem [{\citenamefont {Nayak}\ \emph {et~al.}(2021)\citenamefont {Nayak},
  \citenamefont {Steinbok}, \citenamefont {Roet}, \citenamefont {Koo},
  \citenamefont {Margalit}, \citenamefont {Feldman}, \citenamefont {Almoalem},
  \citenamefont {Kanigel}, \citenamefont {Fiete}, \citenamefont {Yan} \emph
  {et~al.}}]{nayak2021evidence}%
  \BibitemOpen
  \bibfield  {author} {\bibinfo {author} {\bibfnamefont {A.~K.}\ \bibnamefont
  {Nayak}}, \bibinfo {author} {\bibfnamefont {A.}~\bibnamefont {Steinbok}},
  \bibinfo {author} {\bibfnamefont {Y.}~\bibnamefont {Roet}}, \bibinfo {author}
  {\bibfnamefont {J.}~\bibnamefont {Koo}}, \bibinfo {author} {\bibfnamefont
  {G.}~\bibnamefont {Margalit}}, \bibinfo {author} {\bibfnamefont
  {I.}~\bibnamefont {Feldman}}, \bibinfo {author} {\bibfnamefont
  {A.}~\bibnamefont {Almoalem}}, \bibinfo {author} {\bibfnamefont
  {A.}~\bibnamefont {Kanigel}}, \bibinfo {author} {\bibfnamefont {G.~A.}\
  \bibnamefont {Fiete}}, \bibinfo {author} {\bibfnamefont {B.}~\bibnamefont
  {Yan}}, \emph {et~al.},\ }\href@noop {} {\bibfield  {journal} {\bibinfo
  {journal} {Nature physics}\ }\textbf {\bibinfo {volume} {17}},\ \bibinfo
  {pages} {1413} (\bibinfo {year} {2021})}\BibitemShut {NoStop}%
\bibitem [{\citenamefont {Persky}\ \emph {et~al.}(2022)\citenamefont {Persky},
  \citenamefont {Bj{\o}rlig}, \citenamefont {Feldman}, \citenamefont
  {Almoalem}, \citenamefont {Altman}, \citenamefont {Berg}, \citenamefont
  {Kimchi}, \citenamefont {Ruhman}, \citenamefont {Kanigel},\ and\
  \citenamefont {Kalisky}}]{persky2022magnetic}%
  \BibitemOpen
  \bibfield  {author} {\bibinfo {author} {\bibfnamefont {E.}~\bibnamefont
  {Persky}}, \bibinfo {author} {\bibfnamefont {A.~V.}\ \bibnamefont
  {Bj{\o}rlig}}, \bibinfo {author} {\bibfnamefont {I.}~\bibnamefont {Feldman}},
  \bibinfo {author} {\bibfnamefont {A.}~\bibnamefont {Almoalem}}, \bibinfo
  {author} {\bibfnamefont {E.}~\bibnamefont {Altman}}, \bibinfo {author}
  {\bibfnamefont {E.}~\bibnamefont {Berg}}, \bibinfo {author} {\bibfnamefont
  {I.}~\bibnamefont {Kimchi}}, \bibinfo {author} {\bibfnamefont
  {J.}~\bibnamefont {Ruhman}}, \bibinfo {author} {\bibfnamefont
  {A.}~\bibnamefont {Kanigel}},\ and\ \bibinfo {author} {\bibfnamefont
  {B.}~\bibnamefont {Kalisky}},\ }\href@noop {} {\bibfield  {journal} {\bibinfo
   {journal} {Nature}\ }\textbf {\bibinfo {volume} {607}},\ \bibinfo {pages}
  {692} (\bibinfo {year} {2022})}\BibitemShut {NoStop}%
\bibitem [{\citenamefont {Yan}\ \emph {et~al.}(2023)\citenamefont {Yan},
  \citenamefont {Ding}, \citenamefont {Li}, \citenamefont {Tang}, \citenamefont
  {Chen}, \citenamefont {Liu}, \citenamefont {Bu}, \citenamefont {Huang},
  \citenamefont {Dai}, \citenamefont {Jin} \emph {et~al.}}]{yan2023modulating}%
  \BibitemOpen
  \bibfield  {author} {\bibinfo {author} {\bibfnamefont {L.}~\bibnamefont
  {Yan}}, \bibinfo {author} {\bibfnamefont {C.}~\bibnamefont {Ding}}, \bibinfo
  {author} {\bibfnamefont {M.}~\bibnamefont {Li}}, \bibinfo {author}
  {\bibfnamefont {R.}~\bibnamefont {Tang}}, \bibinfo {author} {\bibfnamefont
  {W.}~\bibnamefont {Chen}}, \bibinfo {author} {\bibfnamefont {B.}~\bibnamefont
  {Liu}}, \bibinfo {author} {\bibfnamefont {K.}~\bibnamefont {Bu}}, \bibinfo
  {author} {\bibfnamefont {T.}~\bibnamefont {Huang}}, \bibinfo {author}
  {\bibfnamefont {D.}~\bibnamefont {Dai}}, \bibinfo {author} {\bibfnamefont
  {X.}~\bibnamefont {Jin}}, \emph {et~al.},\ }\href@noop {} {\bibfield
  {journal} {\bibinfo  {journal} {Nano Letters}\ }\textbf {\bibinfo {volume}
  {23}},\ \bibinfo {pages} {2121} (\bibinfo {year} {2023})}\BibitemShut
  {NoStop}%
\bibitem [{\citenamefont {Va{\v{n}}o}\ \emph {et~al.}(2021)\citenamefont
  {Va{\v{n}}o}, \citenamefont {Amini}, \citenamefont {Ganguli}, \citenamefont
  {Chen}, \citenamefont {Lado}, \citenamefont {Kezilebieke},\ and\
  \citenamefont {Liljeroth}}]{vavno2021artificial}%
  \BibitemOpen
  \bibfield  {author} {\bibinfo {author} {\bibfnamefont {V.}~\bibnamefont
  {Va{\v{n}}o}}, \bibinfo {author} {\bibfnamefont {M.}~\bibnamefont {Amini}},
  \bibinfo {author} {\bibfnamefont {S.~C.}\ \bibnamefont {Ganguli}}, \bibinfo
  {author} {\bibfnamefont {G.}~\bibnamefont {Chen}}, \bibinfo {author}
  {\bibfnamefont {J.~L.}\ \bibnamefont {Lado}}, \bibinfo {author}
  {\bibfnamefont {S.}~\bibnamefont {Kezilebieke}},\ and\ \bibinfo {author}
  {\bibfnamefont {P.}~\bibnamefont {Liljeroth}},\ }\href@noop {} {\bibfield
  {journal} {\bibinfo  {journal} {Nature}\ }\textbf {\bibinfo {volume} {599}},\
  \bibinfo {pages} {582} (\bibinfo {year} {2021})}\BibitemShut {NoStop}%
\bibitem [{\citenamefont {Ayani}\ \emph
  {et~al.}(2024{\natexlab{a}})\citenamefont {Ayani}, \citenamefont {Pisarra},
  \citenamefont {Ibarburu}, \citenamefont {Garnica}, \citenamefont {Miranda},
  \citenamefont {Calleja}, \citenamefont {Martín},\ and\ \citenamefont
  {Vázquez~de Parga}}]{ayani2024probing}%
  \BibitemOpen
  \bibfield  {author} {\bibinfo {author} {\bibfnamefont {C.~G.}\ \bibnamefont
  {Ayani}}, \bibinfo {author} {\bibfnamefont {M.}~\bibnamefont {Pisarra}},
  \bibinfo {author} {\bibfnamefont {I.~M.}\ \bibnamefont {Ibarburu}}, \bibinfo
  {author} {\bibfnamefont {M.}~\bibnamefont {Garnica}}, \bibinfo {author}
  {\bibfnamefont {R.}~\bibnamefont {Miranda}}, \bibinfo {author} {\bibfnamefont
  {F.}~\bibnamefont {Calleja}}, \bibinfo {author} {\bibfnamefont
  {F.}~\bibnamefont {Martín}},\ and\ \bibinfo {author} {\bibfnamefont {A.~L.}\
  \bibnamefont {Vázquez~de Parga}},\ }\href
  {https://doi.org/https://doi.org/10.1002/smll.202303275} {\bibfield
  {journal} {\bibinfo  {journal} {Small}\ }\textbf {\bibinfo {volume} {20}},\
  \bibinfo {pages} {2303275} (\bibinfo {year}
  {2024}{\natexlab{a}})}\BibitemShut {NoStop}%
\bibitem [{\citenamefont {Crippa}\ \emph {et~al.}(2024)\citenamefont {Crippa},
  \citenamefont {Bae}, \citenamefont {Wunderlich}, \citenamefont {Mazin},
  \citenamefont {Yan}, \citenamefont {Sangiovanni}, \citenamefont {Wehling},\
  and\ \citenamefont {Valent{\'\i}}}]{crippa2024heavy}%
  \BibitemOpen
  \bibfield  {author} {\bibinfo {author} {\bibfnamefont {L.}~\bibnamefont
  {Crippa}}, \bibinfo {author} {\bibfnamefont {H.}~\bibnamefont {Bae}},
  \bibinfo {author} {\bibfnamefont {P.}~\bibnamefont {Wunderlich}}, \bibinfo
  {author} {\bibfnamefont {I.~I.}\ \bibnamefont {Mazin}}, \bibinfo {author}
  {\bibfnamefont {B.}~\bibnamefont {Yan}}, \bibinfo {author} {\bibfnamefont
  {G.}~\bibnamefont {Sangiovanni}}, \bibinfo {author} {\bibfnamefont
  {T.}~\bibnamefont {Wehling}},\ and\ \bibinfo {author} {\bibfnamefont
  {R.}~\bibnamefont {Valent{\'\i}}},\ }\href@noop {} {\bibfield  {journal}
  {\bibinfo  {journal} {Nature Communications}\ }\textbf {\bibinfo {volume}
  {15}},\ \bibinfo {pages} {1357} (\bibinfo {year} {2024})}\BibitemShut
  {NoStop}%
\bibitem [{\citenamefont {Almoalem}\ \emph {et~al.}(2024)\citenamefont
  {Almoalem}, \citenamefont {Gofman}, \citenamefont {Nitzav}, \citenamefont
  {Mangel}, \citenamefont {Feldman}, \citenamefont {Koo}, \citenamefont
  {Mazzola}, \citenamefont {Fujii}, \citenamefont {Vobornik}, \citenamefont
  {S{\'{}}~anchez Barriga} \emph {et~al.}}]{almoalem2024charge}%
  \BibitemOpen
  \bibfield  {author} {\bibinfo {author} {\bibfnamefont {A.}~\bibnamefont
  {Almoalem}}, \bibinfo {author} {\bibfnamefont {R.}~\bibnamefont {Gofman}},
  \bibinfo {author} {\bibfnamefont {Y.}~\bibnamefont {Nitzav}}, \bibinfo
  {author} {\bibfnamefont {I.}~\bibnamefont {Mangel}}, \bibinfo {author}
  {\bibfnamefont {I.}~\bibnamefont {Feldman}}, \bibinfo {author} {\bibfnamefont
  {J.}~\bibnamefont {Koo}}, \bibinfo {author} {\bibfnamefont {F.}~\bibnamefont
  {Mazzola}}, \bibinfo {author} {\bibfnamefont {J.}~\bibnamefont {Fujii}},
  \bibinfo {author} {\bibfnamefont {I.}~\bibnamefont {Vobornik}}, \bibinfo
  {author} {\bibfnamefont {J.}~\bibnamefont {S{\'{}}~anchez Barriga}}, \emph
  {et~al.},\ }\href@noop {} {\bibfield  {journal} {\bibinfo  {journal} {npj
  Quantum Materials}\ }\textbf {\bibinfo {volume} {9}},\ \bibinfo {pages} {36}
  (\bibinfo {year} {2024})}\BibitemShut {NoStop}%
\bibitem [{\citenamefont {Wen}\ \emph {et~al.}(2021)\citenamefont {Wen},
  \citenamefont {Gao}, \citenamefont {Xie}, \citenamefont {Zhang},
  \citenamefont {Kong}, \citenamefont {Wang}, \citenamefont {Jiang},
  \citenamefont {Luo}, \citenamefont {Li}, \citenamefont {Lu} \emph
  {et~al.}}]{wen2021roles}%
  \BibitemOpen
  \bibfield  {author} {\bibinfo {author} {\bibfnamefont {C.}~\bibnamefont
  {Wen}}, \bibinfo {author} {\bibfnamefont {J.}~\bibnamefont {Gao}}, \bibinfo
  {author} {\bibfnamefont {Y.}~\bibnamefont {Xie}}, \bibinfo {author}
  {\bibfnamefont {Q.}~\bibnamefont {Zhang}}, \bibinfo {author} {\bibfnamefont
  {P.}~\bibnamefont {Kong}}, \bibinfo {author} {\bibfnamefont {J.}~\bibnamefont
  {Wang}}, \bibinfo {author} {\bibfnamefont {Y.}~\bibnamefont {Jiang}},
  \bibinfo {author} {\bibfnamefont {X.}~\bibnamefont {Luo}}, \bibinfo {author}
  {\bibfnamefont {J.}~\bibnamefont {Li}}, \bibinfo {author} {\bibfnamefont
  {W.}~\bibnamefont {Lu}}, \emph {et~al.},\ }\href@noop {} {\bibfield
  {journal} {\bibinfo  {journal} {Physical Review Letters}\ }\textbf {\bibinfo
  {volume} {126}},\ \bibinfo {pages} {256402} (\bibinfo {year}
  {2021})}\BibitemShut {NoStop}%
\bibitem [{\citenamefont {Kumar~Nayak}\ \emph {et~al.}(2023)\citenamefont
  {Kumar~Nayak}, \citenamefont {Steinbok}, \citenamefont {Roet}, \citenamefont
  {Koo}, \citenamefont {Feldman}, \citenamefont {Almoalem}, \citenamefont
  {Kanigel}, \citenamefont {Yan}, \citenamefont {Rosch}, \citenamefont
  {Avraham} \emph {et~al.}}]{kumar2023first}%
  \BibitemOpen
  \bibfield  {author} {\bibinfo {author} {\bibfnamefont {A.}~\bibnamefont
  {Kumar~Nayak}}, \bibinfo {author} {\bibfnamefont {A.}~\bibnamefont
  {Steinbok}}, \bibinfo {author} {\bibfnamefont {Y.}~\bibnamefont {Roet}},
  \bibinfo {author} {\bibfnamefont {J.}~\bibnamefont {Koo}}, \bibinfo {author}
  {\bibfnamefont {I.}~\bibnamefont {Feldman}}, \bibinfo {author} {\bibfnamefont
  {A.}~\bibnamefont {Almoalem}}, \bibinfo {author} {\bibfnamefont
  {A.}~\bibnamefont {Kanigel}}, \bibinfo {author} {\bibfnamefont
  {B.}~\bibnamefont {Yan}}, \bibinfo {author} {\bibfnamefont {A.}~\bibnamefont
  {Rosch}}, \bibinfo {author} {\bibfnamefont {N.}~\bibnamefont {Avraham}},
  \emph {et~al.},\ }\href@noop {} {\bibfield  {journal} {\bibinfo  {journal}
  {Proceedings of the National Academy of Sciences}\ }\textbf {\bibinfo
  {volume} {120}},\ \bibinfo {pages} {e2304274120} (\bibinfo {year}
  {2023})}\BibitemShut {NoStop}%
\bibitem [{\citenamefont {Ritschel}\ \emph {et~al.}(2018)\citenamefont
  {Ritschel}, \citenamefont {Berger},\ and\ \citenamefont
  {Geck}}]{ritschel2018stacking}%
  \BibitemOpen
  \bibfield  {author} {\bibinfo {author} {\bibfnamefont {T.}~\bibnamefont
  {Ritschel}}, \bibinfo {author} {\bibfnamefont {H.}~\bibnamefont {Berger}},\
  and\ \bibinfo {author} {\bibfnamefont {J.}~\bibnamefont {Geck}},\ }\href@noop
  {} {\bibfield  {journal} {\bibinfo  {journal} {Physical Review B}\ }\textbf
  {\bibinfo {volume} {98}},\ \bibinfo {pages} {195134} (\bibinfo {year}
  {2018})}\BibitemShut {NoStop}%
\bibitem [{\citenamefont {Darancet}\ \emph {et~al.}(2014)\citenamefont
  {Darancet}, \citenamefont {Millis},\ and\ \citenamefont
  {Marianetti}}]{darancet2014three}%
  \BibitemOpen
  \bibfield  {author} {\bibinfo {author} {\bibfnamefont {P.}~\bibnamefont
  {Darancet}}, \bibinfo {author} {\bibfnamefont {A.~J.}\ \bibnamefont
  {Millis}},\ and\ \bibinfo {author} {\bibfnamefont {C.~A.}\ \bibnamefont
  {Marianetti}},\ }\href@noop {} {\bibfield  {journal} {\bibinfo  {journal}
  {Physical Review B}\ }\textbf {\bibinfo {volume} {90}},\ \bibinfo {pages}
  {045134} (\bibinfo {year} {2014})}\BibitemShut {NoStop}%
\bibitem [{\citenamefont {Butler}\ \emph {et~al.}(2020)\citenamefont {Butler},
  \citenamefont {Yoshida}, \citenamefont {Hanaguri},\ and\ \citenamefont
  {Iwasa}}]{butler2020mottness}%
  \BibitemOpen
  \bibfield  {author} {\bibinfo {author} {\bibfnamefont {C.}~\bibnamefont
  {Butler}}, \bibinfo {author} {\bibfnamefont {M.}~\bibnamefont {Yoshida}},
  \bibinfo {author} {\bibfnamefont {T.}~\bibnamefont {Hanaguri}},\ and\
  \bibinfo {author} {\bibfnamefont {Y.}~\bibnamefont {Iwasa}},\ }\href@noop {}
  {\bibfield  {journal} {\bibinfo  {journal} {Nature communications}\ }\textbf
  {\bibinfo {volume} {11}},\ \bibinfo {pages} {2477} (\bibinfo {year}
  {2020})}\BibitemShut {NoStop}%
\bibitem [{\citenamefont {Wang}\ \emph {et~al.}(2018)\citenamefont {Wang},
  \citenamefont {Sun}, \citenamefont {Abdelwahab}, \citenamefont {Cao},
  \citenamefont {Yu}, \citenamefont {Ju}, \citenamefont {Zhu}, \citenamefont
  {Fu}, \citenamefont {Chu}, \citenamefont {Xu} \emph
  {et~al.}}]{wang2018surface}%
  \BibitemOpen
  \bibfield  {author} {\bibinfo {author} {\bibfnamefont {Z.}~\bibnamefont
  {Wang}}, \bibinfo {author} {\bibfnamefont {Y.-Y.}\ \bibnamefont {Sun}},
  \bibinfo {author} {\bibfnamefont {I.}~\bibnamefont {Abdelwahab}}, \bibinfo
  {author} {\bibfnamefont {L.}~\bibnamefont {Cao}}, \bibinfo {author}
  {\bibfnamefont {W.}~\bibnamefont {Yu}}, \bibinfo {author} {\bibfnamefont
  {H.}~\bibnamefont {Ju}}, \bibinfo {author} {\bibfnamefont {J.}~\bibnamefont
  {Zhu}}, \bibinfo {author} {\bibfnamefont {W.}~\bibnamefont {Fu}}, \bibinfo
  {author} {\bibfnamefont {L.}~\bibnamefont {Chu}}, \bibinfo {author}
  {\bibfnamefont {H.}~\bibnamefont {Xu}}, \emph {et~al.},\ }\href@noop {}
  {\bibfield  {journal} {\bibinfo  {journal} {ACS nano}\ }\textbf {\bibinfo
  {volume} {12}},\ \bibinfo {pages} {12619} (\bibinfo {year}
  {2018})}\BibitemShut {NoStop}%
\bibitem [{\citenamefont {Sung}\ \emph {et~al.}(2022)\citenamefont {Sung},
  \citenamefont {Schnitzer}, \citenamefont {Novakov}, \citenamefont
  {El~Baggari}, \citenamefont {Luo}, \citenamefont {Gim}, \citenamefont {Vu},
  \citenamefont {Li}, \citenamefont {Brintlinger}, \citenamefont {Liu} \emph
  {et~al.}}]{sung2022two}%
  \BibitemOpen
  \bibfield  {author} {\bibinfo {author} {\bibfnamefont {S.~H.}\ \bibnamefont
  {Sung}}, \bibinfo {author} {\bibfnamefont {N.}~\bibnamefont {Schnitzer}},
  \bibinfo {author} {\bibfnamefont {S.}~\bibnamefont {Novakov}}, \bibinfo
  {author} {\bibfnamefont {I.}~\bibnamefont {El~Baggari}}, \bibinfo {author}
  {\bibfnamefont {X.}~\bibnamefont {Luo}}, \bibinfo {author} {\bibfnamefont
  {J.}~\bibnamefont {Gim}}, \bibinfo {author} {\bibfnamefont {N.~M.}\
  \bibnamefont {Vu}}, \bibinfo {author} {\bibfnamefont {Z.}~\bibnamefont {Li}},
  \bibinfo {author} {\bibfnamefont {T.~H.}\ \bibnamefont {Brintlinger}},
  \bibinfo {author} {\bibfnamefont {Y.}~\bibnamefont {Liu}}, \emph {et~al.},\
  }\href@noop {} {\bibfield  {journal} {\bibinfo  {journal} {Nature
  Communications}\ }\textbf {\bibinfo {volume} {13}},\ \bibinfo {pages} {413}
  (\bibinfo {year} {2022})}\BibitemShut {NoStop}%
\bibitem [{\citenamefont {Sung}\ \emph {et~al.}(2024)\citenamefont {Sung},
  \citenamefont {Agarwal}, \citenamefont {El~Baggari}, \citenamefont {Kezer},
  \citenamefont {Goh}, \citenamefont {Schnitzer}, \citenamefont {Shen},
  \citenamefont {Chiang}, \citenamefont {Liu}, \citenamefont {Lu} \emph
  {et~al.}}]{sung2024endotaxial}%
  \BibitemOpen
  \bibfield  {author} {\bibinfo {author} {\bibfnamefont {S.~H.}\ \bibnamefont
  {Sung}}, \bibinfo {author} {\bibfnamefont {N.}~\bibnamefont {Agarwal}},
  \bibinfo {author} {\bibfnamefont {I.}~\bibnamefont {El~Baggari}}, \bibinfo
  {author} {\bibfnamefont {P.}~\bibnamefont {Kezer}}, \bibinfo {author}
  {\bibfnamefont {Y.~M.}\ \bibnamefont {Goh}}, \bibinfo {author} {\bibfnamefont
  {N.}~\bibnamefont {Schnitzer}}, \bibinfo {author} {\bibfnamefont {J.~M.}\
  \bibnamefont {Shen}}, \bibinfo {author} {\bibfnamefont {T.}~\bibnamefont
  {Chiang}}, \bibinfo {author} {\bibfnamefont {Y.}~\bibnamefont {Liu}},
  \bibinfo {author} {\bibfnamefont {W.}~\bibnamefont {Lu}}, \emph {et~al.},\
  }\href@noop {} {\bibfield  {journal} {\bibinfo  {journal} {Nature
  Communications}\ }\textbf {\bibinfo {volume} {15}},\ \bibinfo {pages} {1403}
  (\bibinfo {year} {2024})}\BibitemShut {NoStop}%
\bibitem [{\citenamefont {Husremovi{\'c}}\ \emph {et~al.}(2023)\citenamefont
  {Husremovi{\'c}}, \citenamefont {Goodge}, \citenamefont {Erodici},
  \citenamefont {Inzani}, \citenamefont {Mier}, \citenamefont {Ribet},
  \citenamefont {Bustillo}, \citenamefont {Taniguchi}, \citenamefont
  {Watanabe}, \citenamefont {Ophus} \emph {et~al.}}]{husremovic2023encoding}%
  \BibitemOpen
  \bibfield  {author} {\bibinfo {author} {\bibfnamefont {S.}~\bibnamefont
  {Husremovi{\'c}}}, \bibinfo {author} {\bibfnamefont {B.~H.}\ \bibnamefont
  {Goodge}}, \bibinfo {author} {\bibfnamefont {M.~P.}\ \bibnamefont {Erodici}},
  \bibinfo {author} {\bibfnamefont {K.}~\bibnamefont {Inzani}}, \bibinfo
  {author} {\bibfnamefont {A.}~\bibnamefont {Mier}}, \bibinfo {author}
  {\bibfnamefont {S.~M.}\ \bibnamefont {Ribet}}, \bibinfo {author}
  {\bibfnamefont {K.~C.}\ \bibnamefont {Bustillo}}, \bibinfo {author}
  {\bibfnamefont {T.}~\bibnamefont {Taniguchi}}, \bibinfo {author}
  {\bibfnamefont {K.}~\bibnamefont {Watanabe}}, \bibinfo {author}
  {\bibfnamefont {C.}~\bibnamefont {Ophus}}, \emph {et~al.},\ }\href@noop {}
  {\bibfield  {journal} {\bibinfo  {journal} {Nature Communications}\ }\textbf
  {\bibinfo {volume} {14}},\ \bibinfo {pages} {6031} (\bibinfo {year}
  {2023})}\BibitemShut {NoStop}%
\bibitem [{\citenamefont {Ravnik}\ \emph {et~al.}(2019)\citenamefont {Ravnik},
  \citenamefont {Vaskivskyi}, \citenamefont {Gerasimenko}, \citenamefont
  {Diego}, \citenamefont {Vodeb}, \citenamefont {Kabanov},\ and\ \citenamefont
  {Mihailovic}}]{ravnik2019strain}%
  \BibitemOpen
  \bibfield  {author} {\bibinfo {author} {\bibfnamefont {J.}~\bibnamefont
  {Ravnik}}, \bibinfo {author} {\bibfnamefont {I.}~\bibnamefont {Vaskivskyi}},
  \bibinfo {author} {\bibfnamefont {Y.}~\bibnamefont {Gerasimenko}}, \bibinfo
  {author} {\bibfnamefont {M.}~\bibnamefont {Diego}}, \bibinfo {author}
  {\bibfnamefont {J.}~\bibnamefont {Vodeb}}, \bibinfo {author} {\bibfnamefont
  {V.}~\bibnamefont {Kabanov}},\ and\ \bibinfo {author} {\bibfnamefont {D.~D.}\
  \bibnamefont {Mihailovic}},\ }\href@noop {} {\bibfield  {journal} {\bibinfo
  {journal} {ACS applied nano materials}\ }\textbf {\bibinfo {volume} {2}},\
  \bibinfo {pages} {3743} (\bibinfo {year} {2019})}\BibitemShut {NoStop}%
\bibitem [{\citenamefont {Ravnik}\ \emph {et~al.}(2021)\citenamefont {Ravnik},
  \citenamefont {Vaskivskyi}, \citenamefont {Vodeb}, \citenamefont
  {Aupi{\v{c}}}, \citenamefont {Vaskivskyi}, \citenamefont {Gole{\v{z}}},
  \citenamefont {Gerasimenko}, \citenamefont {Kabanov},\ and\ \citenamefont
  {Mihailovic}}]{ravnik2021quantum}%
  \BibitemOpen
  \bibfield  {author} {\bibinfo {author} {\bibfnamefont {J.}~\bibnamefont
  {Ravnik}}, \bibinfo {author} {\bibfnamefont {Y.}~\bibnamefont {Vaskivskyi}},
  \bibinfo {author} {\bibfnamefont {J.}~\bibnamefont {Vodeb}}, \bibinfo
  {author} {\bibfnamefont {P.}~\bibnamefont {Aupi{\v{c}}}}, \bibinfo {author}
  {\bibfnamefont {I.}~\bibnamefont {Vaskivskyi}}, \bibinfo {author}
  {\bibfnamefont {D.}~\bibnamefont {Gole{\v{z}}}}, \bibinfo {author}
  {\bibfnamefont {Y.}~\bibnamefont {Gerasimenko}}, \bibinfo {author}
  {\bibfnamefont {V.}~\bibnamefont {Kabanov}},\ and\ \bibinfo {author}
  {\bibfnamefont {D.}~\bibnamefont {Mihailovic}},\ }\href@noop {} {\bibfield
  {journal} {\bibinfo  {journal} {Nature Communications}\ }\textbf {\bibinfo
  {volume} {12}},\ \bibinfo {pages} {3793} (\bibinfo {year}
  {2021})}\BibitemShut {NoStop}%
\bibitem [{\citenamefont {Kresse}\ and\ \citenamefont
  {Joubert}(1999)}]{kresse1999ultrasoft}%
  \BibitemOpen
  \bibfield  {author} {\bibinfo {author} {\bibfnamefont {G.}~\bibnamefont
  {Kresse}}\ and\ \bibinfo {author} {\bibfnamefont {D.}~\bibnamefont
  {Joubert}},\ }\href@noop {} {\bibfield  {journal} {\bibinfo  {journal}
  {Physical Review B}\ }\textbf {\bibinfo {volume} {59}},\ \bibinfo {pages}
  {1758} (\bibinfo {year} {1999})}\BibitemShut {NoStop}%
\bibitem [{\citenamefont {Perdew}\ \emph {et~al.}(1996)\citenamefont {Perdew},
  \citenamefont {Burke},\ and\ \citenamefont
  {Ernzerhof}}]{perdew1996generalized}%
  \BibitemOpen
  \bibfield  {author} {\bibinfo {author} {\bibfnamefont {J.~P.}\ \bibnamefont
  {Perdew}}, \bibinfo {author} {\bibfnamefont {K.}~\bibnamefont {Burke}},\ and\
  \bibinfo {author} {\bibfnamefont {M.}~\bibnamefont {Ernzerhof}},\ }\href@noop
  {} {\bibfield  {journal} {\bibinfo  {journal} {Physical Review Letters}\
  }\textbf {\bibinfo {volume} {77}},\ \bibinfo {pages} {3865} (\bibinfo {year}
  {1996})}\BibitemShut {NoStop}%
\bibitem [{\citenamefont {Grimme}\ \emph {et~al.}(2011)\citenamefont {Grimme},
  \citenamefont {Ehrlich},\ and\ \citenamefont {Goerigk}}]{grimme2011effect}%
  \BibitemOpen
  \bibfield  {author} {\bibinfo {author} {\bibfnamefont {S.}~\bibnamefont
  {Grimme}}, \bibinfo {author} {\bibfnamefont {S.}~\bibnamefont {Ehrlich}},\
  and\ \bibinfo {author} {\bibfnamefont {L.}~\bibnamefont {Goerigk}},\
  }\href@noop {} {\bibfield  {journal} {\bibinfo  {journal} {Journal of
  computational chemistry}\ }\textbf {\bibinfo {volume} {32}},\ \bibinfo
  {pages} {1456} (\bibinfo {year} {2011})}\BibitemShut {NoStop}%
\bibitem [{\citenamefont {Dudarev}\ \emph {et~al.}(1998)\citenamefont
  {Dudarev}, \citenamefont {Botton}, \citenamefont {Savrasov}, \citenamefont
  {Humphreys},\ and\ \citenamefont {Sutton}}]{dudarev1998electron}%
  \BibitemOpen
  \bibfield  {author} {\bibinfo {author} {\bibfnamefont {S.~L.}\ \bibnamefont
  {Dudarev}}, \bibinfo {author} {\bibfnamefont {G.~A.}\ \bibnamefont {Botton}},
  \bibinfo {author} {\bibfnamefont {S.~Y.}\ \bibnamefont {Savrasov}}, \bibinfo
  {author} {\bibfnamefont {C.}~\bibnamefont {Humphreys}},\ and\ \bibinfo
  {author} {\bibfnamefont {A.~P.}\ \bibnamefont {Sutton}},\ }\href@noop {}
  {\bibfield  {journal} {\bibinfo  {journal} {Physical Review B}\ }\textbf
  {\bibinfo {volume} {57}},\ \bibinfo {pages} {1505} (\bibinfo {year}
  {1998})}\BibitemShut {NoStop}%
\bibitem [{\citenamefont {Ayani}\ \emph
  {et~al.}(2024{\natexlab{b}})\citenamefont {Ayani}, \citenamefont {Bosnar},
  \citenamefont {Calleja}, \citenamefont {Sol{\'e}}, \citenamefont
  {Stetsovych}, \citenamefont {Ibarburu}, \citenamefont {Rebanal},
  \citenamefont {Garnica}, \citenamefont {Miranda}, \citenamefont {Otrokov}
  \emph {et~al.}}]{ayani2024unveiling}%
  \BibitemOpen
  \bibfield  {author} {\bibinfo {author} {\bibfnamefont {C.~G.}\ \bibnamefont
  {Ayani}}, \bibinfo {author} {\bibfnamefont {M.}~\bibnamefont {Bosnar}},
  \bibinfo {author} {\bibfnamefont {F.}~\bibnamefont {Calleja}}, \bibinfo
  {author} {\bibfnamefont {A.~P.}\ \bibnamefont {Sol{\'e}}}, \bibinfo {author}
  {\bibfnamefont {O.}~\bibnamefont {Stetsovych}}, \bibinfo {author}
  {\bibfnamefont {I.~M.}\ \bibnamefont {Ibarburu}}, \bibinfo {author}
  {\bibfnamefont {C.}~\bibnamefont {Rebanal}}, \bibinfo {author} {\bibfnamefont
  {M.}~\bibnamefont {Garnica}}, \bibinfo {author} {\bibfnamefont
  {R.}~\bibnamefont {Miranda}}, \bibinfo {author} {\bibfnamefont {M.~M.}\
  \bibnamefont {Otrokov}}, \emph {et~al.},\ }\href@noop {} {\bibfield
  {journal} {\bibinfo  {journal} {Nano Letters}\ }\textbf {\bibinfo {volume}
  {24}},\ \bibinfo {pages} {10805} (\bibinfo {year}
  {2024}{\natexlab{b}})}\BibitemShut {NoStop}%
\bibitem [{\citenamefont {Fazekas}\ and\ \citenamefont
  {Tosatti}(1979)}]{fazekas1979electrical}%
  \BibitemOpen
  \bibfield  {author} {\bibinfo {author} {\bibfnamefont {P.}~\bibnamefont
  {Fazekas}}\ and\ \bibinfo {author} {\bibfnamefont {E.}~\bibnamefont
  {Tosatti}},\ }\href@noop {} {\bibfield  {journal} {\bibinfo  {journal}
  {Philosophical Magazine B}\ }\textbf {\bibinfo {volume} {39}},\ \bibinfo
  {pages} {229} (\bibinfo {year} {1979})}\BibitemShut {NoStop}%
\bibitem [{\citenamefont {Lee}\ \emph {et~al.}(2019)\citenamefont {Lee},
  \citenamefont {Goh},\ and\ \citenamefont {Cho}}]{lee2019origin}%
  \BibitemOpen
  \bibfield  {author} {\bibinfo {author} {\bibfnamefont {S.-H.}\ \bibnamefont
  {Lee}}, \bibinfo {author} {\bibfnamefont {J.~S.}\ \bibnamefont {Goh}},\ and\
  \bibinfo {author} {\bibfnamefont {D.}~\bibnamefont {Cho}},\ }\href@noop {}
  {\bibfield  {journal} {\bibinfo  {journal} {Physical review letters}\
  }\textbf {\bibinfo {volume} {122}},\ \bibinfo {pages} {106404} (\bibinfo
  {year} {2019})}\BibitemShut {NoStop}%
\bibitem [{\citenamefont {Ritschel}\ \emph {et~al.}(2015)\citenamefont
  {Ritschel}, \citenamefont {Trinckauf}, \citenamefont {Koepernik},
  \citenamefont {B{\"u}chner}, \citenamefont {Zimmermann}, \citenamefont
  {Berger}, \citenamefont {Joe}, \citenamefont {Abbamonte},\ and\ \citenamefont
  {Geck}}]{ritschel2015orbital}%
  \BibitemOpen
  \bibfield  {author} {\bibinfo {author} {\bibfnamefont {T.}~\bibnamefont
  {Ritschel}}, \bibinfo {author} {\bibfnamefont {J.}~\bibnamefont {Trinckauf}},
  \bibinfo {author} {\bibfnamefont {K.}~\bibnamefont {Koepernik}}, \bibinfo
  {author} {\bibfnamefont {B.}~\bibnamefont {B{\"u}chner}}, \bibinfo {author}
  {\bibfnamefont {M.~v.}\ \bibnamefont {Zimmermann}}, \bibinfo {author}
  {\bibfnamefont {H.}~\bibnamefont {Berger}}, \bibinfo {author} {\bibfnamefont
  {Y.}~\bibnamefont {Joe}}, \bibinfo {author} {\bibfnamefont {P.}~\bibnamefont
  {Abbamonte}},\ and\ \bibinfo {author} {\bibfnamefont {J.}~\bibnamefont
  {Geck}},\ }\href@noop {} {\bibfield  {journal} {\bibinfo  {journal} {Nature
  physics}\ }\textbf {\bibinfo {volume} {11}},\ \bibinfo {pages} {328}
  (\bibinfo {year} {2015})}\BibitemShut {NoStop}%
\bibitem [{\citenamefont {Pizarro}\ \emph {et~al.}(2020)\citenamefont
  {Pizarro}, \citenamefont {Adler}, \citenamefont {Zantout}, \citenamefont
  {Mertz}, \citenamefont {Barone}, \citenamefont {Valent{\'\i}}, \citenamefont
  {Sangiovanni},\ and\ \citenamefont {Wehling}}]{pizarro2020deconfinement}%
  \BibitemOpen
  \bibfield  {author} {\bibinfo {author} {\bibfnamefont {J.~M.}\ \bibnamefont
  {Pizarro}}, \bibinfo {author} {\bibfnamefont {S.}~\bibnamefont {Adler}},
  \bibinfo {author} {\bibfnamefont {K.}~\bibnamefont {Zantout}}, \bibinfo
  {author} {\bibfnamefont {T.}~\bibnamefont {Mertz}}, \bibinfo {author}
  {\bibfnamefont {P.}~\bibnamefont {Barone}}, \bibinfo {author} {\bibfnamefont
  {R.}~\bibnamefont {Valent{\'\i}}}, \bibinfo {author} {\bibfnamefont
  {G.}~\bibnamefont {Sangiovanni}},\ and\ \bibinfo {author} {\bibfnamefont
  {T.~O.}\ \bibnamefont {Wehling}},\ }\href@noop {} {\bibfield  {journal}
  {\bibinfo  {journal} {npj quantum materials}\ }\textbf {\bibinfo {volume}
  {5}},\ \bibinfo {pages} {79} (\bibinfo {year} {2020})}\BibitemShut {NoStop}%
\bibitem [{\citenamefont {Parr}\ and\ \citenamefont
  {Pearson}(1983)}]{parr1983absolute}%
  \BibitemOpen
  \bibfield  {author} {\bibinfo {author} {\bibfnamefont {R.~G.}\ \bibnamefont
  {Parr}}\ and\ \bibinfo {author} {\bibfnamefont {R.~G.}\ \bibnamefont
  {Pearson}},\ }\href@noop {} {\bibfield  {journal} {\bibinfo  {journal}
  {Journal of the American chemical society}\ }\textbf {\bibinfo {volume}
  {105}},\ \bibinfo {pages} {7512} (\bibinfo {year} {1983})}\BibitemShut
  {NoStop}%
\bibitem [{\citenamefont {Read}\ and\ \citenamefont {Green}(2000)}]{Read2000}%
  \BibitemOpen
  \bibfield  {author} {\bibinfo {author} {\bibfnamefont {N.}~\bibnamefont
  {Read}}\ and\ \bibinfo {author} {\bibfnamefont {D.}~\bibnamefont {Green}},\
  }\href {https://doi.org/10.1103/PhysRevB.61.10267} {\bibfield  {journal}
  {\bibinfo  {journal} {Phys. Rev. B}\ }\textbf {\bibinfo {volume} {61}},\
  \bibinfo {pages} {10267} (\bibinfo {year} {2000})}\BibitemShut {NoStop}%
\bibitem [{\citenamefont {Ivanov}(2001)}]{Ivanov2001}%
  \BibitemOpen
  \bibfield  {author} {\bibinfo {author} {\bibfnamefont {D.~A.}\ \bibnamefont
  {Ivanov}},\ }\href {https://doi.org/10.1103/PhysRevLett.86.268} {\bibfield
  {journal} {\bibinfo  {journal} {Phys. Rev. Lett.}\ }\textbf {\bibinfo
  {volume} {86}},\ \bibinfo {pages} {268} (\bibinfo {year} {2001})}\BibitemShut
  {NoStop}%
\bibitem [{\citenamefont {Duckheim}\ and\ \citenamefont
  {Brouwer}(2011)}]{Duckheim2011}%
  \BibitemOpen
  \bibfield  {author} {\bibinfo {author} {\bibfnamefont {M.}~\bibnamefont
  {Duckheim}}\ and\ \bibinfo {author} {\bibfnamefont {P.~W.}\ \bibnamefont
  {Brouwer}},\ }\href {https://doi.org/10.1103/PhysRevB.83.054513} {\bibfield
  {journal} {\bibinfo  {journal} {Phys. Rev. B}\ }\textbf {\bibinfo {volume}
  {83}},\ \bibinfo {pages} {054513} (\bibinfo {year} {2011})}\BibitemShut
  {NoStop}%
\bibitem [{\citenamefont {Chung}\ \emph {et~al.}(2011)\citenamefont {Chung},
  \citenamefont {Zhang}, \citenamefont {Qi},\ and\ \citenamefont
  {Zhang}}]{Chung2011}%
  \BibitemOpen
  \bibfield  {author} {\bibinfo {author} {\bibfnamefont {S.~B.}\ \bibnamefont
  {Chung}}, \bibinfo {author} {\bibfnamefont {H.-J.}\ \bibnamefont {Zhang}},
  \bibinfo {author} {\bibfnamefont {X.-L.}\ \bibnamefont {Qi}},\ and\ \bibinfo
  {author} {\bibfnamefont {S.-C.}\ \bibnamefont {Zhang}},\ }\href
  {https://doi.org/10.1103/PhysRevB.84.060510} {\bibfield  {journal} {\bibinfo
  {journal} {Phys. Rev. B}\ }\textbf {\bibinfo {volume} {84}},\ \bibinfo
  {pages} {060510} (\bibinfo {year} {2011})}\BibitemShut {NoStop}%
\end{thebibliography}

\end{document}